\documentclass[11pt,twoside,nohyper]{pallis}
\usepackage{epsfig}
\usepackage{latexsym}
\usepackage{amssymb}
\usepackage{fancyhed}

\newcommand{\beq}{\begin{equation}}
\newcommand{\eeq}{\end{equation}}
\newcommand{\ben}{\begin{equation*}}
\newcommand{\een}{\end{equation*}}
\newcommand{\ba}{\begin{eqnarray}}
\newcommand{\ea}{\end{eqnarray}}
\newcommand{\ban}{\begin{eqnarray*}}
\newcommand{\ean}{\end{eqnarray*}}
\newcommand{\brr}{\begin{array}}
\newcommand{\err}{\end{array}}
\newcommand{\bc}{\begin{center}}
\newcommand{\ec}{\end{center}}

\newcommand{\bea}{\begin{eqnarray}}
\newcommand{\eea}{\end{eqnarray}}
\newcommand{\bean}{\begin{eqnarray*}}
\newcommand{\eean}{\end{eqnarray*}}
\newcommand{\ftn}{\footnotesize}
\newcommand{\nsz}{\normalsize}
\newcommand{\ssz}{\scriptsize}
\newcommand{\tr}{{\mbox{\sf\ssz T}}}

 \newcommand{\stau}{\mbox{$\tilde\tau_2$}}
 \newcommand{\staus}{\mbox{$\tilde\tau^\ast_2$}}

\newcommand{\sbb}{\mbox{$\tilde b_2$}}
\newcommand{\sbbs}{\mbox{$\tilde b^\ast_2$}}

\newcommand{\ntn}{\mbox{$\tilde\nu_\tau$}}
\newcommand{\ntns}{\mbox{$\tilde\nu^\ast_\tau$}}

\newcommand{\nta}{\mbox{$\tilde\chi$}}
\newcommand{\ntb}{\mbox{${\tilde\chi_2^0}$}}

\newcommand{\chb}{\mbox{${\tilde \chi_2^\pm}$}}

\newcommand{\chbp}{\mbox{${\tilde \chi_2^+}$}}
\newcommand{\chbm}{\mbox{${\tilde \chi_2^-}$}}

\JPub{Nucl. Phys. B678, 398 (2004)}

\preprint{\tt SISSA-27/2003/EP }

\title{\huge {\Large\boldmath $b-\tau$} U{\kern-1.5pt\Large NIFICATION}
{\Large WITH} G{\kern-1.5pt\Large AUGINO \\ AND} S{\Large FERMION}
M{\Large ASS NON}-U\kern-1.5pt\Large NIVERSALITY}

\author{\Large C. P\kern-1.5pt\nsz ALLIS\\
SISSA/ISAS, Via Beirut 2-4, 34013 Trieste, ITALY\\
\email{pallis@sissa.it}}

\abstract{In the context of a SUSY GUT inspired MSSM version, the
low energy consequences of the asymptotic $b-\tau$ Yukawa coupling
Unification are examined, under the assumption of universal or
non-universal boundary conditions for the gaugino and sfermion
masses. Gaugino non-universality is applied, so that the SUSY
corrections to $b$-quark mass can be reconciled with the present
experimental data on muon anomalous magnetic moment. Restrictions
on the parameter space, originating from the cold dark matter
abundance in the universe, the inclusive branching ratio of
$b\rightarrow s\gamma$ and the accelerator data are, also,
investigated and the scalar neutralino-proton cross section is
calculated. In the case of a bino-like LSP and universal boundary
conditions for the sfermion masses, the constraints, arising from
the cold dark matter and ${\rm BR}(b \to s\gamma)$ can be
simultaneously satisfied, mainly thanks to the $A$-pole effect or
the neutralino-stau coannihilations. In addition, sfermion mass
non-universality provides the possibility of new coannihilation
phenomena (neutralino-sbottom or neutralino-tau sneutrino-stau),
which facilitate the simultaneous satisfaction of all the other
requirements. In both cases above, the neutralino abundance can
essentially decrease for a W-ino or higgsino like LSP creating
regions of parameter space with additional neutralino-chargino
and/or heavier neutralino coannihilations. The neutralino-sbottom
mass proximity significantly ameliorates the detectability of LSP.
}

\begin{document}

\setcounter{page}{1} \pagestyle{fancyplain}

\addtolength{\headheight}{.5cm}

\rhead[\fancyplain{}{ \bf \thepage}]{\fancyplain{}{ $b-\tau$
U{\ftn NIFICATION WITH} G{\ssz AUGINO AND} S{\ftn FERMION} M{\ftn
ASS NON}-U{\ftn NIVERSALITY}}} \lhead[\fancyplain{}{
\leftmark}]{\fancyplain{}{\bf \thepage}} \cfoot{}

\section{I{\ftn NTRODUCTION}}\label{sec:intro}

\hspace{.562cm} In some recent papers \cite{subir, balj}, the
early prediction in the context of $SU(5)$ Grand Unified Theory
(GUT) \cite{Lang} of the asymptotic $b-\tau$ Yukawa coupling
Unification (YU) was beautifully combined with the present
experimental data on neutrino physics within $SO(10)$ models.
However, applying this scheme in the framework of the Constrained
Minimal Supersymmetric (SUSY) Standard Model (CMSSM) \cite{Cmssm}
and given the top and tau experimental masses, the $\mu$ parameter
is restricted to negative values \cite{kazakov, komine}. This is
due to the fact that the tree level $b$-quark mass receives
sizeable SUSY corrections \cite{copw}, which can drive the
corrected $b$-quark mass within its experimental range only for
$\mu<0$. On the other hand, the $\mu<0$ case is in conflict with
the present experimental data \cite{muon} on the muon anomalous
magnetic moment \cite{kazakov, komine}. Indeed, the deviation,
$\delta a_\mu$, of the measured value of the muon anomalous
magnetic moment from its predicted value in the Standard Model
(SM), seems to favor the $\mu>0$ regime \cite{davier, narison}. In
addition, the negative sign of $\mu$ is severely restricted by the
recent experimental results \cite{cleo} on the inclusive branching
ratio ${\rm BR}(b\rightarrow s\gamma)$ \cite{kazakov, komine},
which bounds below the SUSY spectrum to rather high values.

At the same time, the SUSY spectrum can be bounded above by the
requirement that the relic density, $\Omega_{\rm LSP}h^2$ of the
lightest SUSY particle (LSP) in the universe does not exceed the
upper limit on the Cold Dark Matter (CDM) abundance implied by
cosmological considerations. After the recent experimental results
of WMAP \cite{wmap}, this limit turns out to be as much stringent
as the previous one (see, e.g. Ref. \cite{dasi}) with a
significantly better accuracy (see sec. \ref{phenoa}). Since
$\Omega_{\rm LSP}h^2$ increases with the LSP mass $m_{\rm LSP}$,
this limit imposes a very strong upper bound on $m_{\rm LSP}$.
However, this can significantly weaken in regions of the parameter
space, where a substantial reduction of $\Omega_{\rm LSP}h^2$ can
be achieved, mainly thanks to the $A$ pole effect ($A$PE) and/or
the coannihilation mechanism (CAM). In the first case, which is
applicable for large $\tan\beta$ \cite{lah, cmssm}, the
$\Omega_{\rm LSP}h^2$ reduction is caused by the presence of a
resonance ($2m_{\rm LSP}\simeq m_{\rm A}$) in the $A$ mediated
LSP' s annihilation channel. On the other hand, CAM is activated
for any $\tan\beta$, when a mass proximity occurs between the LSP
and the next to LSP (NLSP). In the context of CMSSM, NLSP can be
the lightest slepton \cite{ellis1, ellis2} and particularly stau
\cite{cdm, cmssm} for large $\tan\beta$ or stop \cite{santoso1,
boem} for very large trilinear coupling, or also, chargino
\cite{baer, darkn}. Possible non-universality in the Higgs sector
\cite{ellis3} gives rise to CAMs between LSP and
$e$/$\mu$-sneutrino and/or chargino-sleptons.

As induced from the previous considerations, the viability of the
$b-\tau$ YU hypothesis in the context of CMSSM becomes rather
dubious. On the other hand, the embedding of MSSM into a $SU(5)$
or $SO(10)$ SUSY GUT leads, also, to a variety of possibilities
\cite{impact} beyond the CMSSM universality. In this paper we will
employ these scenaria in order to obtain SUSY spectra compatible
with all these Cosmo-Phenomenological requirements.  Namely non
universal gaugino masses (UGMs) \cite{nath1, nath2} are applied,
so that the inconsistency between the $\delta a_\mu$ constraint
and the $b$-quark mass experimental data is removed. This effect
can produce gaugino inspired CAMs (neutralino-chargino and/or
heavier neutralino, $\nta-\chb-\ntb$). On the other hand, new
sfermionic CAMs (neutralino-sbottom, $\nta - \sbb$  and
neutralino-tau sneutrino-stau, $\nta -\ntn - \stau$) can be caused
by applying non universal sfermion masses (USFMs) \cite{santoso,
baery}. Both phenomena by themselves or in conjunction can
drastically reduce $\Omega_{\rm LSP}h^2$ to an acceptably low
level, increasing the upper bound on $m_{\rm LSP}$ almost up to
$2~{\rm TeV}$. As a bonus, in both latter cases, the satisfaction
of the BR($b\to s\gamma$) criterion is facilitated, allowing
viable parameter space consistent even with the optimistic upper
bound on $m_{\rm LSP}$ from $\delta a_\mu$ constraint (see sec.
\ref{phenoc}). Consequently the neutralino-proton cross sections
are sensitively increased, especially in the case of $\nta - \sbb$
CAM.





Combination of $b-\tau$ YU and non UGMs were previously considered
in Refs. \cite{nath1, nath2}, where, contrary to our case, a
single dominant direction in gauge kinetic function has been
assumed (see sec. \ref{sec:model}). Additional presence of non
USFMs were not studied until now, while the $\nta -\sbb$ CAM was
just noticed in Ref. \cite{santoso}. Our main improvements are the
consideration of the $\delta a_\mu$ constraint, the reproduction
of the $\nta - \sbb$ CAM in a much more restrictive framework, the
finding of the $\nta -\ntn - \stau$ CAM and the study of the
co-existence of the gaugino inspired CAMs.


The framework of our analysis is described in some detail in sec.
\ref{sec:pheno}. The basic features of our model are established
in sec. \ref{sec:model}. Our numerical approach and results are
exhibited in secs. \ref{num} and  \ref{result}. We end up with our
conclusions and some open issues in sec. \ref{con}. Throughout the
text and the formulas, brackets are used by applying disjunctive
correspondence.


\section{C{\ftn OSMO}-P{\ftn HENOMENOLOGICAL}
F{\ftn RAMEWORK}} \label{sec:pheno}

\hspace{.562cm}  We briefly describe the operation of the various
Cosmo-Phenomenological criteria that we will use in our
investigation. In the following formulas, gaugino masses, $M_2,
M_3$, top trilinear coupling $A_t$, $\mu$ parameter and the
various SUSY corrections, $\Delta m_{b[\tau]}$ to $b$-quark
[$\tau$-lepton] mass are calculated, at a SUSY breaking scale,
$M_{\rm SUSY}$ specified in sec. \ref{num}.

\subsection{C{\ssz OLD} D{\ssz ARK} M{\ssz ATTER} C{\ssz ONSIDERATIONS}}
\label{phenoa}

\hspace{.562cm} According to WMAP results \cite{wmap}, the total
(M) and the baryonic (B) matter abundance in the universe is
respectively:
\beq \Omega_{\rm
M}h^2=0.135_{-0.009}^{+0.008}~~\mbox{and}~~\Omega_{\rm
B}h^2=0.0224\pm0.0009. \eeq
We, thus, deduce the $95\%$ confidence level (c.l.) range for the
CDM abundance \cite{spanos}:
\beq \Omega_{\rm CDM}h^2=0.1126_{-0.0181}^{+0.0161}.
\label{cdmba}\eeq

In the context of MSSM, the lightest neutralino, $\tilde\chi$ can
be the LSP. It consists the most natural candidate for solving the
CDM problem, being neutral, weakly interacting and stable in the
context of SUSY theories with $R$-Parity \cite{goldberg}
conservation. Hence, in our analysis, we require the LSP relic
density not to exceed the upper bound of Eq. (\ref{cdmba}):
\beq \Omega_{\rm LSP}h^2\lesssim0.13.\label{cdmb}\eeq

We calculate $\Omega_{\rm LSP}h^2$, using {\tt micrOMEGAs}
\cite{micro}, which is one of the most complete publicly available
codes. This includes accurately thermally averaged exact
tree-level cross sections of all possible (co)annihilation
processes, treats poles properly and uses one loop QCD corrections
to the Higgs decay widths and couplings into fermions
\cite{width}.

\subsection{S{\ssz CALAR} N{\ssz EUTRALINO}-P{\ssz ROTON} C{\ssz ROSS} S{\ssz ECTION}}
\label{phenoaa}

\hspace{.562cm} Neutralinos could be detected via their elastic
scattering with nuclei \cite{early}. The quantity which is being
conventionally used in the literature (e.g. \cite{efo, drees1}) to
compare experimental \cite{exp, expp} and theoretical results is
the scalar neutralino-proton ($\nta-p$) cross section,
\begin{eqnarray}
\sigma_{\tilde\chi p}^{\rm SI}=4\,\mu^2_{\tilde\chi
p}f_p^2/\pi~~{\rm where}~~\mu_{\tilde\chi p}=m_{\rm
LSP}m_p/(m_{\rm LSP}+m_p)
\end{eqnarray}
and $f_p$ is the scalar contribution to the effective $\nta-p$
coupling. We calculate it, using the full one loop treatment of
Ref. \cite{drees1} (with some typos fixed \cite{jungman}). This is
indispensable for a reliable result in the region $m_{\rm LSP}\sim
m_{\tilde b_2}$, where the tree level approximation (which, indeed
works well elsewhere) fails. For the involved
renormalization-invariant functions were adopted the values
\cite{efo} (in the notation of Ref. \cite{drees1}):
\begin{eqnarray}
&& f_{{\rm T}_u}^p=0.02\pm0.004,\qquad
   f_{{\rm T}_d}^p=0.026\pm0.005,\qquad
   f_{{\rm T}_s}^p=0.118\pm0.062. \label{rgis}\end{eqnarray}
Combining the sensitivities of the recent \cite{exp} and planned
\cite{expp} experiments, we end in the following
phenomenologically interesting region, for $100~{\rm GeV}\lesssim
m_{\rm LSP}\lesssim 500~{\rm GeV}$:
\beq \mbox{\sf a)}~~3\times 10^{-9}~{\rm pb}\lesssim
\sigma_{\tilde\chi p}^{\rm SI}~~~~\mbox{and}~~~~
\mbox{\sf b)}~~\sigma_{\tilde\chi p}^{\rm SI}\lesssim 2 \times
10^{-6}~{\rm pb} \label{sgmb} \eeq
For SUSY spectra of our models consistent with all the other
constraints of sec. \ref{sec:pheno}, the obtained
$\sigma_{\tilde\chi p}^{\rm SI}$ lies beyond the claimed by
{\small DAMA} preferred range, $(1-10)\times 10^{-6}~\rm{pb}$,
which however, has mostly been excluded by other collaborations
(e.g. {\small EDELWEISS, ZEPLIN} I).



\subsection{SUSY C{\ssz  ORRECTIONS TO} $b$-Q{\ssz UARK AND}
$\tau$-L{\ssz EPTON} M{\ssz ASS}} \label{phenob}

\hspace{.562cm} In the large and intermediate $\tan\beta$ regime,
the tree level $b$-quark mass, $m_b$ receives sizeable SUSY
corrections \cite{copw, pierce, susy}, $\Delta m_b$ which arise
from sbottom-gluino, $\left(\Delta m_b\right)^{\tilde b\tilde g}$
(mainly) and stop-chargino, $(\Delta m_b)^{\tilde
t\tilde\chi^\pm}$ loops. $(\Delta m_b)^{\tilde t\tilde\chi^\pm}$
interferes destructively (see Eq. \ref{bsgs})) to $\left(\Delta
m_b\right)^{\tilde b\tilde g}$ which, normally (an exception is
constructed in Ref. \cite{raby}) dominates $\Delta m_b$.
Consequently:
\beq {\rm sign}~\Delta m_b={\rm sign}~
M_3\>\mu,~~\mbox{since}~~{\rm sign}~\left(\Delta
m_b\right)^{\tilde b\tilde g[\tilde t\tilde\chi^\pm]}={\rm sign}~
M_3~[A_t]\>\mu,\label{mbs} \eeq
using the standard sign convention of Ref. \cite{sugra}. Hence,
for $M_3\mu>[<]~0$, the corrected $b$-quark mass at a low scale
$M_Z$,
\beq m^{\rm c}_b(M_Z)=m_b(M_Z)\left(1+\Delta m_b\right),
\label{mbrn} \eeq
turns out to be above [below] its tree level value $m_b(M_Z)$. The
result is to be compared with the $95\%$ c.l. experimental range
for $m^{\rm c}_b(M_Z)$. This is derived by appropriately
\cite{qcdm} evolving the corresponding range for the pole
$b$-quark mass, $m_b(m_b)$ up to $M_Z$ scale with
$\alpha_s(M_Z)\simeq 0.1185$, in accord with the analysis in Ref.
\cite{baermb}:
\beq m_b(m_b) = 4.25\pm 0.3\>{\rm
GeV}\>\>\>{\Longrightarrow}\>\>\>m^{\rm c}_b(M_Z)=2.88\pm
0.2\>{\rm GeV}. \label{mbb}\eeq

Less important but not negligible (almost 5$\%$) are the SUSY
corrections to $\tau$-lepton, $\Delta m_\tau$ originated from
\cite{pierce} sneutrino-chargino, $\left(\Delta
m_\tau\right)^{\tilde \nu_\tau\tilde\chi^\pm}$ (mainly) and
stau-neutralino, $(\Delta m_\tau)^{\tilde \tau\tilde\chi}$ loops,
with the following signs:
\beq {\rm sign}~\Delta m_\tau=-{\rm sign}~
M_2\>\mu,~~\mbox{since}~~{\rm sign}~\left(\Delta
m_\tau\right)^{\tilde \nu_\tau\tilde\chi^\pm [\tilde
\tau\tilde\chi]}=-{\rm sign}~ M_2~[-M_1]\>\mu.\label{mtaus} \eeq

\subsection{B{\ssz RANCHING} R{\ssz ATIO OF}
$b\rightarrow s\gamma$}\label{phenog}

\hspace{.562cm} Taking into account the recent experimental
results \cite{cleo} on this ratio, ${\rm BR}(b\rightarrow
s\gamma)$, and combining appropriately the experimental and
theoretical involved errors \cite{qcdm}, we obtain the following
$95\%$ c.l. range:
\beq \mbox{\sf a)}~~1.9\times 10^{-4}\lesssim {\rm
BR}(b\rightarrow s\gamma)~~~~\mbox{and}~~~~
\mbox{\sf b)}~~{\rm BR}(b\rightarrow s\gamma)\lesssim 4.6 \times
10^{-4} \label{bsgb} \eeq
We compute ${\rm BR}(b\rightarrow s\gamma)$ by using an updated
version of the relevant calculation contained in the {\tt
micrOMEGAs} package \cite{micro}. In this code, the SM
contribution is calculated using the formalism of Ref.
\cite{kagan} including the improvements of Ref. \cite{gambino}.
The $H^\pm$ contribution is evaluated by including the
next-to-leading order (NLO) QCD corrections from Ref.
\cite{nlohiggs} and $\tan\beta$ enhanced contributions from Ref.
\cite{nlosusy}. The dominant SUSY contribution, ${\rm
BR}(b\rightarrow s\gamma)|_{\rm SUSY}$, includes resummed NLO SUSY
QCD corrections from Ref. \cite{nlosusy}, which hold for large
$\tan\beta$. The $H^\pm$ contribution interferes constructively
with the SM contribution, whereas ${\rm BR}(b\rightarrow
s\gamma)|_{\rm SUSY}$ interferes de[con]-structively with the
other two contributions for $-M_3\mu<[>]~0$, since
\cite{borzumati}, in general:
\beq {\rm sign}~{\rm BR}(b\rightarrow s\gamma)|_{\rm SUSY}={\rm
sign}~A_t\>\mu,~~\mbox{with}~~{\rm sign}~A_t=-{\rm sign}~M_3.
\label{bsgs} \eeq
%
However, the SM plus $H^\pm$ contributions and the ${\rm
BR}(b\rightarrow s\gamma)|_{\rm SUSY}$ decrease as $m_{\rm LSP}$
increases and so, a lower bound on $m_{\rm LSP}$ can be derived
from Eq. (\ref{bsgb}{\sf a}~[\ref{bsgb}{\sf b}]) for
$M_3\mu>[<]~0$ with the latter being much more restrictive. It is
obvious from Eqs. (\ref{bsgs}) and (\ref{mbs}) that simultaneous
combination of negative correction to $b$-quark mass and
destructive contribution of ${\rm BR}(b\rightarrow s\gamma)|_{\rm
SUSY}$ is impossible \cite{borzumati}.

\subsection{M{\ssz UON} A{\ssz  NOMALOUS}
M{\ssz  AGNETIC} M{\ssz  OMENT}}\label{phenoc}

\hspace{.562cm} The deviation, $\delta a_\mu$ of the $a_\mu$
measured value from its predicted value in the SM, $a^{\rm
SM}_\mu$ can be attributed to SUSY contributions, arising from
chargino-sneutrino and neutralino-smuon loops. $\delta a_\mu$ is
calculated by using {\tt micrOMEGAs} routine based on the
formul\ae of Ref. \cite{gmuon}. The absolute value of the result
decreases as $m_{\rm LSP}$ increases and its sign is:
\beq {\rm sign}~\delta a_\mu={\rm sign}~ M_2\>\mu.\label{g2s} \eeq
On the other hand, the $a^{\rm SM}_\mu$ calculation is not yet
stabilized mainly due to the instability of the hadronic vacuum
polarization contribution. According to the evaluation of this
contribution in Ref. \cite{davier}, the findings based on $e^+e^-$
annihilation and $\tau$-decay data are inconsistent with each
other. Taking into account these results and the recently
announced experimental measurements \cite{muon} on $a_\mu$, we
impose the following $95\%$ c.l. ranges:
\bea \mbox{\sf a)}\hspace{16.95pt}11.3\times10^{-10}\lesssim\delta
a_\mu&~~\mbox{and}~~&\mbox{\sf b)}~~\delta a_\mu\lesssim
56.1\times 10^{-10}~~~~\mbox{$e^+e^-$-based}\label{g2e}
\\\vspace*{19pt}
\mbox{\sf a)}~-11.6\times10^{-10}\lesssim\delta
a_\mu&~~\mbox{and}~~&\mbox{\sf b)}~~\delta a_\mu\lesssim30.4\times
10^{-10}~~~~\mbox{$\tau$-based}\label{g2t} \eea
A lower bound on $m_{\rm LSP}$ can be derived for $M_2\mu>[<]~0$
from Eq. (\ref{g2e}{\sf b} [\ref{g2t}{\sf a}]) and an optimistic
upper bound for $M_2\mu>0$ from Eq. (\ref{g2e}{\sf a}) which,
however is not imposed as an absolute constraint due to the former
computational instabilities. Although the $M_2\mu<0$ case can not
be excluded \cite{gmuon2}, it is considered as quite disfavored
\cite{narison}, because of the poor $\tau$-decay data. For this
reason, following the common practice \cite{ellis3, spanos}, we
adopt the restrictions to parameters induced from Eq. (\ref{g2e}).

\subsection{C{\ssz OLLIDER} B{\ssz OUNDS}}
\label{phenod}

\hspace{.562cm} The relevant for our analysis is the $95\%$ c.l.
LEP bound on the lightest CP-even neutral Higgs boson
\cite{higgs}, $h$ and the lightest sbottom \cite{sbott}, $\sbb$
mass,
\beq\mbox{\sf a)}~~m_h\gtrsim114.4~{\rm GeV}~~
\mbox{and}~~\mbox{\sf b)}~~m_{\tilde b_2} \gtrsim95~{\rm GeV}.
\label{mhb}\eeq
The SUSY corrections to $m_h$ are calculated at two-loop by using
the {\tt FeynHiggsFast} \cite{fh} program included in {\tt
micrOMEGAs} code \cite{micro}.

\section{P{\ftn ARTICLE} M{\ftn ODEL}}\label{sec:model}

\hspace{.562cm} The embedding of MSSM in a SUSY GUT enriches the
model with extra constraints and opens new possibilities beyond
the CMSSM universality \cite{impact}. We below select some of
them, constructing the Yukawa (sec. \ref{sec:yuk}), gaugino (sec.
\ref{sec:gau}) and scalar (sec. \ref{sec:sca}) sector of our
theory. Actually, it is a variant of the models proposed in secs.
II and III of Ref. \cite{impact}. The phenomenological reasons
which push us in the introduction of the non-universalities of
sec. \ref{sec:gau} and \ref{sec:sca} will become obvious during
the presentation of our results in sec. \ref{result}. However, for
clarity, let us outline them shortly. Consistency of $b-\tau$ YU
with Eq. (\ref{g2e}{\sf a}) requires a proper application of non
UGMs. The resulting model has still two shortcomings: (i)
Uninteresting $\sigma_{\tilde\chi p}$ due to large minimal $m_{\rm
LSP}$, since this is derived from Eq. (\ref{bsgb}{\sf b}) (ii)
Inability for the satisfaction of Eq. (\ref{g2e}{\sf b}). Non
USFMs assist us to alleviate both disadvantages.

\subsection{$b-\tau$ U{\ssz NIFICATION}}\label{sec:yuk}

\hspace{.562cm} In the minimal $SU(5)$ SUSY GUT \cite{impact,
komine}, the third generation left handed superfields
$L=(\nu_{\tau},\tau)$, $b^c$ belong to the ${\bf \bar5}$
representation (reps), while $Q=(t,b)$, $t^c$, $\tau^c$ belong to
the ${\bf 10}$ reps. Assuming that the electroweak Higgs
superfields $H_1$, $H_2$ are contained in ${\bf\bar 5}_H$ and
${\bf 5}_H$ reps, respectively, the model predicts $b-\tau$ YU at
GUT scale, $M_{\rm GUT}$ ($M_{\rm GUT}$ is determined by the
requirement of gauge coupling unification):
\beq h_b(M_{\rm GUT})=h_\tau(M_{\rm GUT})=y_{b\tau} \label{exact}
\eeq
since, the Yukawa coupling terms of the resulting version of MSSM:
\beq h_b H_1^\tr i \tau_2 Q\; b^c +h_\tau H_1^\tr i\tau_2 L\;
\tau^c\label{ffy} \eeq
originate from an unique term, $y_{b\tau}\ {\bf \bar5}\ {\bf 10}\
{\bf \bar5}_H$ of the underlying GUT.

The asymptotic relation of Eq. (\ref{exact}) can also arise in the
context of $SO(10)$ SUSY GUT. In this case, a family of fermions
is incorporated in the ${\bf  16}$ spinorial reps.  Assuming that
$H_1$, $H_2$ are contained in two different Higgses \cite{impact,
subir} in ${\bf 10}_{H_D}$ and ${\bf 10}_{H_U}$, the terms in Eq.
(\ref{ffy}) can be again derived from an unique term, $y_{b\tau}\
{\bf 16}\ {\bf 16}\ {\bf 10}_{H_D}$. Alternatively, if $H_1$,
$H_2$ are expressed as a combination of ${\bf 10}_{H}$ and ${\bf
\overline{126}}_{H}$, large atmospheric neutrino mixing, produced
through a non canonical see-saw mechanism, requires \cite{balj}
$b-\tau$ YU.

Assuming exact $b-\tau$ YU at $M_{\rm GUT}$ and given the top and
tau masses, $\tan\beta$ and $m^{\rm c}_b(M_Z)$ can not be both
free parameters (see, also, sec. \ref{num}). We choose as input
parameter in our presentation $m^{\rm c}_b(M_Z)$ (in contrast with
the usual in the literature strategy, see, e.g. Refs.
\cite{kazakov, komine, nath1, nath2}). As a consequence, a
prediction can be made for $\tan\beta$. Furthermore, the sign of
$\Delta m_b$ has to be negative. This is, because close to the
complete YU ($\tan\beta\;\simeq 50$), we obtain $m_b(M_Z)$ close
to the upper edge of the range of Eq. (\ref{mbb}). As $\tan\beta$
decreases, $m_b(M_Z)$ increases \cite{anant} and so, a negative
$\Delta m_b$ can drive $m^{\rm c}_b(M_Z)$ for some values of
$\tan\beta<50$ (see, e.g. Fig. 4 of Ref. \cite{hw}) within the
above range. Combining this result with Eq. (\ref{mbs}), we
conclude that $b-\tau$ YU can become viable only for $M_3\mu<0$
(in accordance with the findings of Refs. \cite{kazakov, komine,
nath1, nath2}).

\subsection{G{\ssz AUGINO} S{\ssz ECTOR}}\label{sec:gau}

\hspace{.562cm} From the discussion of the previous section, we
can induce that $b-\tau$ YU in the context of CMSSM \cite{Cmssm},
is viable only for $\mu<0$. Consequently the parameter space of
the model can be restricted through Eq. (\ref{g2t}), which,
however, is rather oracular. To liberate the model from this ugly
feature, we invoke a departure from the UGMs. The importance of
non UGMs in addressing the former inconsistency has already been
stressed in Refs. \cite{nath1, nath2}. Indeed, from Eqs.
(\ref{mbs}) and (\ref{g2s}), we can infer that negativity of
$\Delta m_b$ and positivity of $\delta a_\mu$, can be reconciled
with the following arrangement:
\beq M_1(M_{\rm GUT}):M_2(M_{\rm GUT}):M_3(M_{\rm GUT}) =
1:+[-]r_2:-[+]r_3,~~ \mbox{with}~~ \mu>[<]~0,\label{ratio}\eeq
and $r_2, r_3>0$. Such a condition can arise, by employing a
moderate deviation from the minimal Supergravity (mSUGRA) scenario
\cite{eent, andbaer, kawam, corsetti, bno} (for an other approach
see, e.g. Refs. \cite{dermi}) as follows.

According to the gravity-mediated SUSY breaking mechanism
\cite{martin}, gaugino masses $M_i, i=1,2,3$ are generated by a
left handed chiral superfield $\Phi$, which appears linearly in
the gauge kinetic function $f_{\alpha\beta}$ ($\alpha,\beta$ run
over the GUT gauge group generators). During the spontaneous
breaking of the GUT symmetry, auxiliary fields $F_{\Phi}$,
components of $\Phi$, coupled to gauginos, acquire vacuum
expectation values (vevs), which can be considered as the
asymptotic $M_i$. In mSUGRA models, the fields $F_{\Phi}$ are
treated as singlets under the underlying GUT and therefore, UGMs
result. However, $F_{\Phi}$ can belong to any reps $r$ in the
symmetric product ($\rm S$) of two adjoints \cite{eent}. Thus,
after the breaking of the GUT symmetry to the SM one, $F^r_{\Phi}$
acquire vevs in the SM neutral direction $\langle
F^r_{\Phi}\rangle_{\alpha\beta} =n^r_\alpha\delta_{\alpha\beta}$,
where $n^r_\alpha$ are group theoretical factors. Therefore, the
gaugino masses at $M_{\rm GUT}$ can be parameterized as follows
($i<\alpha$):
\beq  M_i(M_{\rm GUT})=M_{1/2}\sum_r c_r n_i^r, \label{genr} \eeq
where $M_{1/2}$ is a gaugino mass parameter and the characteristic
numbers $n_i^r$ of every reps $r$ have been worked out in Ref.
\cite{anderson} for the $SU(5)$ and in Ref. \cite{so10g} for the
$SO(10)$ GUT. Also, $c_r$ are the relative weight of the reps $r$
in the sum. They can be treated as free parameters, while there is
no direct experimental constraint on the resulting signs
\cite{kawam}. As emphasized in Refs. \cite{andbaer, impact}, all
the cases are compatible with the gauge coupling unification with
the assumption that the scalar component of $\Phi$ develops
negligible vev. One can next show, by solving the resulting 3x3
systems (we ignore for simplicity the large reps {\bf 220} [{\bf
770}] for $SU(5)~[SO(10)]$), that there is a wide and natural set
of $c_r$ 's in Eq. (\ref{genr}) for $0<r_{2,3}<2$, so that the
ratio of Eq. (\ref{ratio}) can be realized. To keep our
investigation as general as possible, we will not assume (as
usually \cite{nath1, andbaer, so10g}) dominance of a specific
direction in Eq. (\ref{genr}). Nevertheless, we will comment on
these more restrictive but certainly more predictive cases, in the
conclusions.

We will close this section, quoting several important comments:

{\bf i.} In both cases of Eq. (\ref{ratio}), ${\rm
BR}(b\rightarrow s\gamma)|_{\rm SUSY}$ interferes constructively
to SM+$H^\pm$ contribution and therefore, the satisfaction of Eq.
(\ref{bsgb}{\sf b}) has to be attained.

{\bf ii.} The renormalization group running is affected very
little by the specific choice one of the two possibilities in Eq.
(\ref{ratio}). However, we choose to work with $\mu>0$, since in
this case the resulting value of $\tan\beta$ is slightly
diminished. This is due to the fact that for $\mu>0$, $\Delta
m_\tau$  increases (due to the additive correlation of the two
contributions in Eq. (\ref{mtaus}) which does not exist for
$\mu<0$ with $M_2<0$ from Eq. (\ref{ratio})). As a consequence,
with given the tau yukawa coupling $h_\tau(M_{\rm SUSY})$, a
larger $m_\tau(M_{\rm SUSY})$ (or smaller $\tan\beta$) for $\mu>0$
is needed so as a successful $m_\tau(M_Z)$ is obtained (see, also,
sec. \ref{num}).

{\bf iii.} For $r_2>1$ and $r_3>1$, LSP is mainly a pure B-ino.
However for $r_2<1$ and/or $r_3<1$, LSP can become W-ino or
Higgsino like and the mass of charginos and/or gluinos decrease so
as they can coannihilate with LSP reducing $\Omega_{\rm LSP}h^2$
even lower than the expectations \cite{nelson}. Despite the fact
that $\Omega_{\rm LSP}h^2$ much lower than the bound of Eq.
(\ref{cdmba}) can not be characterized as a fatal disadvantage of
the theory (since other production mechanisms of LSP may be
activated \cite{snc} and/or other CDM candidates may contribute to
$\Omega_{\rm CDM}h^2$) we will keep our investigation in regions
of $r_2, r_3$, where $\Omega_{\rm LSP}h^2$ turns out to be close
to the bound of Eq. (\ref{cdmba}) for $m_{\rm LSP}<2~{\rm TeV}$.

\subsection{S{\ssz CALAR} S{\ssz ECTOR}}\label{sec:sca}

\hspace{.562cm} In the minimal $SU(5)$ SUSY GUT \cite{pomarol,
impact}, the soft SUSY breaking terms for the sfermions in ${\bf
10}$ reps ($\tilde Q=(\tilde t,\tilde b)$, $\tilde t^c$, $\tilde
\tau^c$), $m_{\bf 10}$ and in ${\bf \bar5}$ reps ($\tilde
L=(\tilde \nu_{\tau},\tilde \tau)$, $\tilde b^c$), $m_{\bf \bar5}$
can be different. Consequently, the soft SUSY breaking masses for
the sfermions of the resulting MSSM at $M_{\rm GUT}$ can be
written as:
\begin{eqnarray}
m_{\tilde Q}^2(M_{\rm GUT})=m_{\tilde t^c}^2(M_{\rm
GUT})=m_{\tilde \tau^c}^2(M_{\rm GUT}) &=& m_0^2, \label{da}\\
m_{\tilde L}^2(M_{\rm GUT})=m_{\tilde b^c}^2(M_{\rm GUT}) &=&
r^2_{\tilde f}m_0^2,~~0<r^2_{\tilde f}<2 \label{de}
\end{eqnarray}
where we have adopted the parameterization and the range for
$r^2_{\tilde f}$ of Refs. \cite{sant, santoso}. Similar splitting
between sfermion masses can also occur in the context of $SO(10)$
\cite{dterms, impact, baery} although in the presence of some
$D$-terms. To suppress dangerous flavor changing neutral currents
\cite{masiero}, we maintain the universality among generations.
Remarkably, such a non-universality among sleptons can be probed
at $e^+e^-$ colliders \cite{slbaer}. As we will see in sec.
\ref{result}, $r^2_{\tilde f}<1$ can decrease considerably the
resulting masses of $\sbb$ and $\ntn$ so as to have a chance to
play the r\^ole of coannihilator during the LSP freeze out in the
Early Universe, reducing efficiently $\Omega_{\rm LSP}h^2$.

As regards the soft masses of the two higgses $H_1,H_2$, included
in ${\bf\bar 5}_H~[{\bf 5}_H]$, they can be in general arbitrary
(even in the case of $SO(10)$ SUSY GUT since we assumed two
higgses ${\bf 10}_{H_U}, {\bf 10}_{H_D}$ in sec. \ref{sec:yuk})
\begin{eqnarray}
m_{H_1}^2(M_{\rm GUT})=r^2_{H_1} m_0^2
~~\mbox{and}~~m_{H_2}^2(M_{\rm GUT}) =r^2_{H_2} m_0^2,
~~0<r^2_{H_1,H_2}<2\label{dh}
\end{eqnarray}
However, trying to isolate the non USFMs and reduce the number of
the free parameters, we will restrict ourselves to the
simplificative case $r_{H_1}=r_{H_2}=1$. Nevertheless, we will
comment on the conclusion how this assumption can influence our
results.

\section{N{\ftn UMERICAL} C{\ftn ALCULATION} }\label{num}

\hspace{.562cm} In our numerical calculation, we closely follow
the notation as well as the renormalization group and radiative
electroweak symmetry breaking (RESB) analysis of Refs.
\cite{qcdm}. We integrate the 2-loop renormalization group
equations (RGEs) for the gauge and Yukawa coupling constants and
1-loop for the soft SUSY breaking terms between $M_{\rm GUT}$ and
a common SUSY threshold $M_{\rm SUSY} \simeq(m_{\tilde
t_1}m_{\tilde t_2})^{1/2}$ ($\tilde t_{1,2}$ are the stop mass
eigenstates) determined in consistency with the SUSY spectrum. At
$M_{\rm SUSY}$ we impose RESB, evaluate the SUSY spectrum and
incorporate the SUSY corrections to $b$ and $\tau$ masses
\cite{pierce, susy}. Between $M_{\rm SUSY}$ and $M_Z$, the running
of gauge and Yukawa couplings is continued using the SM RGEs.

We use fixed values for the running top quark mass
$m_t(m_t)=166~{\rm GeV}$ and tau lepton mass
$m_\tau(M_Z)=1.746~{\rm{GeV}}$. Using an iterative up-down
approach, $h_t(M_{\rm GUT})$ and $h_\tau(M_{\rm GUT})$ are
determined for each $\tan\beta$ at $M_{\rm SUSY}$, while
$h_b(M_{\rm GUT})$ is derived from the $b-\tau$ YU assumption, Eq.
(\ref{exact}). Equivalently, turning the procedure around,
$\tan\beta$ can be adjusted so, that the derived $h_b(M_{\rm
GUT})$ corresponds to a desired $m^{\rm c}_b(M_Z)$. Fixing it,
also, to its central experimental value of Eq. (\ref{mbb}),
$m^{\rm c}_b(M_Z)=2.88~{\rm GeV}$, a prediction for $\tan\beta$
can be made, as already mentioned in sec. \ref{sec:yuk}. Finally,
we impose the boundary conditions given by Eqs. (\ref{ratio}) with
$\mu>0$ for the gaugino masses, by Eqs. (\ref{da})-(\ref{dh}) for
the scalar masses and we also, assume a universal trilinear scalar
coupling, $A_0$.

In summary, our effective theory below $M_{\rm GUT}$ depends on
the parameters :
$$M_{1/2},~m_0,~A_0,~r_2,~r_3,~r_{\tilde f}.$$
To further reduce the parameter space of the model, we fix (as
usually \cite{cmssm, ellis3, spanos}) $A_0=0$. $A_0\neq0$ is not
expected to change dramatically our results. Also, for
presentation purposes, $M_{1/2}$ and $m_0$ can be replaced by
$m_{\rm LSP}$ and a relative mass splitting, $\Delta_{\rm P}$,
defined as follows:
\beq \Delta_{\rm P}=\left\{\matrix{
%
(m_{\rm P}- 2m_{\rm LSP })/2m_{\rm LSP}\hfill , & \mbox{if}~~ {\rm
P}:A \hfill \cr
(m_{\rm P} -m_{\rm LSP })/m_{\rm LSP}\hfill ,  & \mbox{if}~~ {\rm
P}:\stau, \sbb~\mbox{or}~\ntn \hfill \cr}
\right. \label{delta} \eeq
The choice of this parameter is convenient, since it determines,
for given $m_{\rm LSP }$, the strength of the $A$PE for P : $A$ or
of the CAM for P : $\stau, \sbb~\mbox{and}~\ntn$. It, thus,
essentially unifies the description of both reduction
``procedures'' (see sec. \ref{result}). Note that although
$\Delta_{\tilde\chi^\pm_2}~[\Delta_{\tilde\chi^0_2}]$ can be
defined through a relation similar to this in the second line of
Eq. (\ref{delta}) with P : $\tilde\chi^\pm_2~[\tilde\chi^0_2]$,
these can not be used in order to determine the spectrum, since
they depend crucially only on $M_{1/2}$ and not on $m_0$. So, they
vary very slowly, once $r_2$ or $r_3$ have been chosen.
Consequently, our final set of the considered free parameters is:
$$m_{\rm LSP},\ \Delta_{\rm P},\ r_2,\ r_3,\ r_{\tilde f}.$$

\section{R{\ssz ESULTS}}\label{result}

\hspace{.562cm} We proceed, now in the delineation of the
parameter space of our model. For the sake of illustration, we
divide this section in subsections devoted to each applied
$\Omega_{\rm LSP}h^2$ reduction ``procedure''. The CAMs are
classified to 3 main categories based to the sfermionic ones with
or without the presence of $\nta-\chb-\ntb$ CAMs. In this case,
the allowed ranges of the basic parameters $m_{\rm LSP},
\Delta_{\rm P}$ and $\tan\beta$ are listed comparatively in the
Tables 1, 2, 3 together with the relative contributions beyond a
threshold value of the (co)annihilation processes to the
$\Omega_{\rm LSP}h^2$ calculation as $m_{\rm LSP}$ and
$\Delta_{\tilde\tau_2}$ vary in their allowed (or, indicative in
some cases) ranges. The allowed regions on the $m_{\rm
LSP}-\Delta_{\rm P}$ plane from the various absolute constraints
of sec. \ref{sec:pheno} are shaded, while the regions favored by
the optimistic upper bound of Eq. (\ref{g2e}{\sf a}) are ruled.
For simplicity, we do not show bounds from constraints less
restrictive than those which are crucial.

Let us introductionary explain the reasons for which we will focus
on some specific $r_2, r_3$ and $r_{\tilde f}$. Initially, in
order to make contact with the highly predictive and
well-investigated parameter space of CMSSM (see, e.g. Refs.
\cite{santoso1, baer, darkn, ellis2, spanos}) we will consider
$r_2=r_3=r_{\tilde f}=1$. Indeed, for $r_2=r_3=1$, the resulting
low energy values of the soft SUSY breaking terms turn out to be
quite similar to those that we would have obtained, if we had
imposed UGMs with $\mu<0$. The gaugino running (see, e.g. Ref.
\cite{rge}) and essentially, the LSP gaugino purity, $G_P$ (in the
notation of Ref. \cite{rge}), remain unaltered. The scalar running
is altered by a few percent due to the resulting lower values of
the trilinear couplings. This is, because their running crucially
depends on the relative sign of $M_2$ and $M_3$. The latter
difference has the following remarkable consequences. In our case:
(i) $\Delta m_b$ turns out to be larger. This is because,
$\left(\Delta m_b\right)^{\tilde t\tilde\chi^\pm}$ anti-correlates
more weakly with $\left(\Delta m_b\right)^{\tilde b\tilde g}$ due
to lower $|A_t|$ (sec. \ref{phenob}). (ii) $\tan\beta$ is
significantly decreased (especially in the case of universal
sfermion masses), since the tree level $m_b(M_Z)$ has to be larger
(sec. \ref{sec:yuk}), so that after the subtraction of the larger
$\Delta m_b$, the resulting $m^{\rm c}_b(M_Z)$ is within its
experimental margin of Eq. (\ref{mbb}). (iii) ${\rm
BR}(b\rightarrow s\gamma)$ is lower, since ${\rm BR}(b\rightarrow
s\gamma)|_{\rm SUSY}$ and the $H^\pm$ contribution is diminished,
mainly due \cite{nlosusy} to the larger denominator of the
resummation and lower $\tan\beta$ enhanced contributions,
respectively.

For $r_2=r_3=1$ and $r_{\tilde f}=1, 0.2, 0.4$, we will present
the mass parameters and the allowed regions on the $m_{\rm
LSP}-\Delta_{\rm P}$ plane. Possible variation of $r_2, r_3$ is
not expected to change the general characteristics of the mass
parameters. Also, we checked that $r_2>1$ and/or $r_3>1$ do not
create new CAMs and so, do not essentially deform the allowed
regions. However, for $r_2<1$ and/or $r_3<1$, additional CAMs can
further enlarge them. A first example will be given for $r_2=0.6$
and $r_3=1$. With this choice, $\Delta_{\tilde\chi^\pm_2}\sim 0.1$
is established, creating a background of useful (not very drastic)
$\nta-\chb-\ntb$ CAMs (in accord with Ref. \cite{nelson}) which
can be combined with the sfemionic ones. However, essential
reduction of $G_P$ is obtained only for $r_3<1$. For $r_2=1$ and
$r_3<1$, new situation for the $\Omega_{\rm LSP}h^2$ calculation
emerges for $r_3<0.5$. Then, $\Delta_{\tilde\chi^\pm_2}< 0.1$ and
$\nta-\chb-\ntb$ CAMs reduce $\Omega_{\rm LSP}h^2$ lower than the
expectations. Instead, we will insist on the choice, $r_2=0.6$ and
$r_3=0.5$ or 0.6, for which $\nta-\chb-\ntb$ CAMs can be kept
under control.

Finally, let the hadronic inputs of Eq. (\ref{rgis}) vary within
their ranges we will derive the corresponding bands on the $m_{\rm
LSP}-\sigma_{\tilde\chi p}^{\rm SI}$ plane for various
$\Delta_{\rm P}$ and $r_2=r_3=r_{\tilde f}=1$, while possible
improvement for $r_2<1$ and $r_3<1$ will be illustrated, too. The
findings will be collectively presented in Fig. \ref{csec}, but
the explanations will be given separately in each subsection.

\begin{figure}[t]
\hspace*{-.71in}
\begin{minipage}{8in}
\epsfig{file=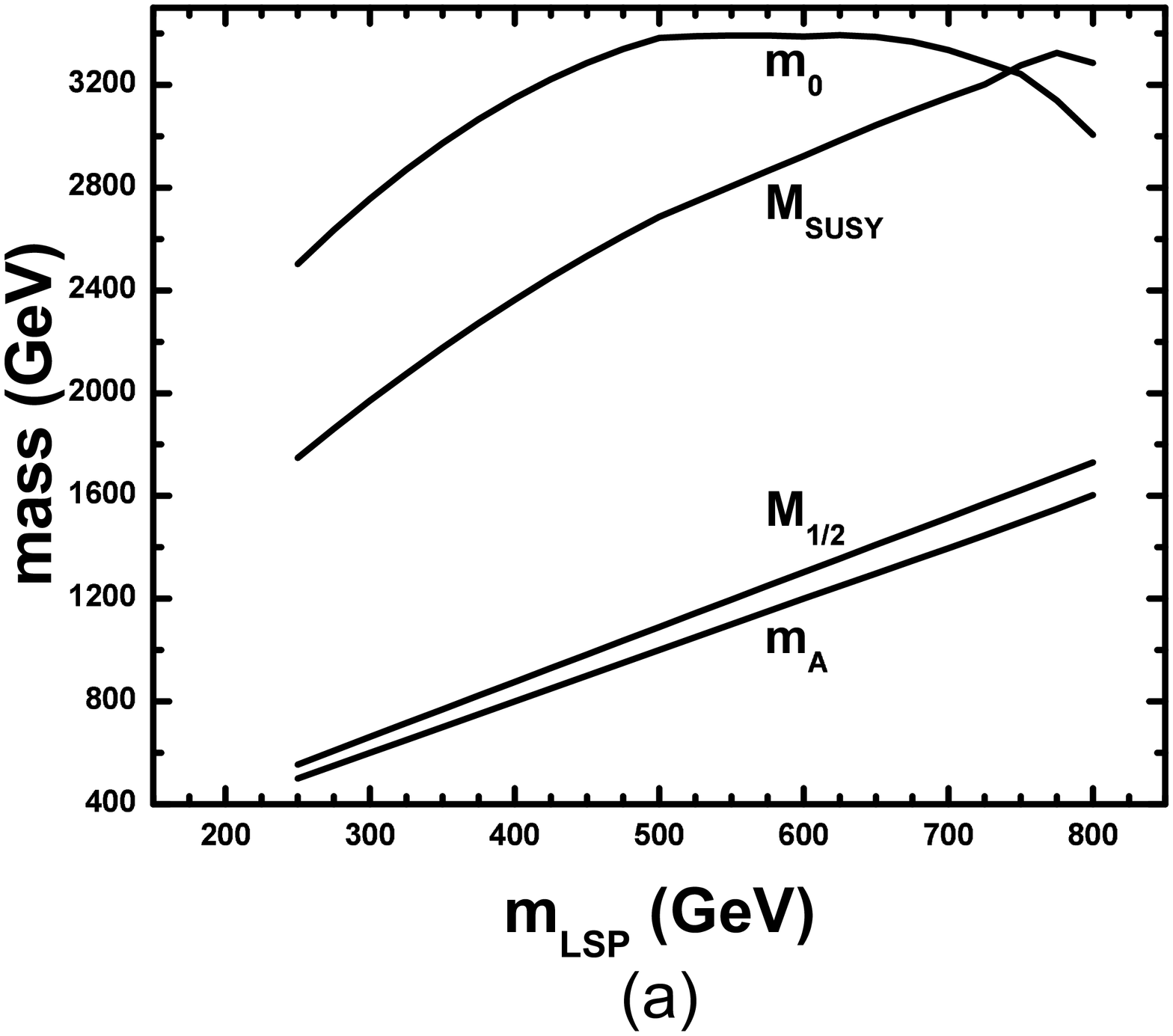,height=3.8in,angle=-90} \hspace*{-1.37 cm}
\epsfig{file=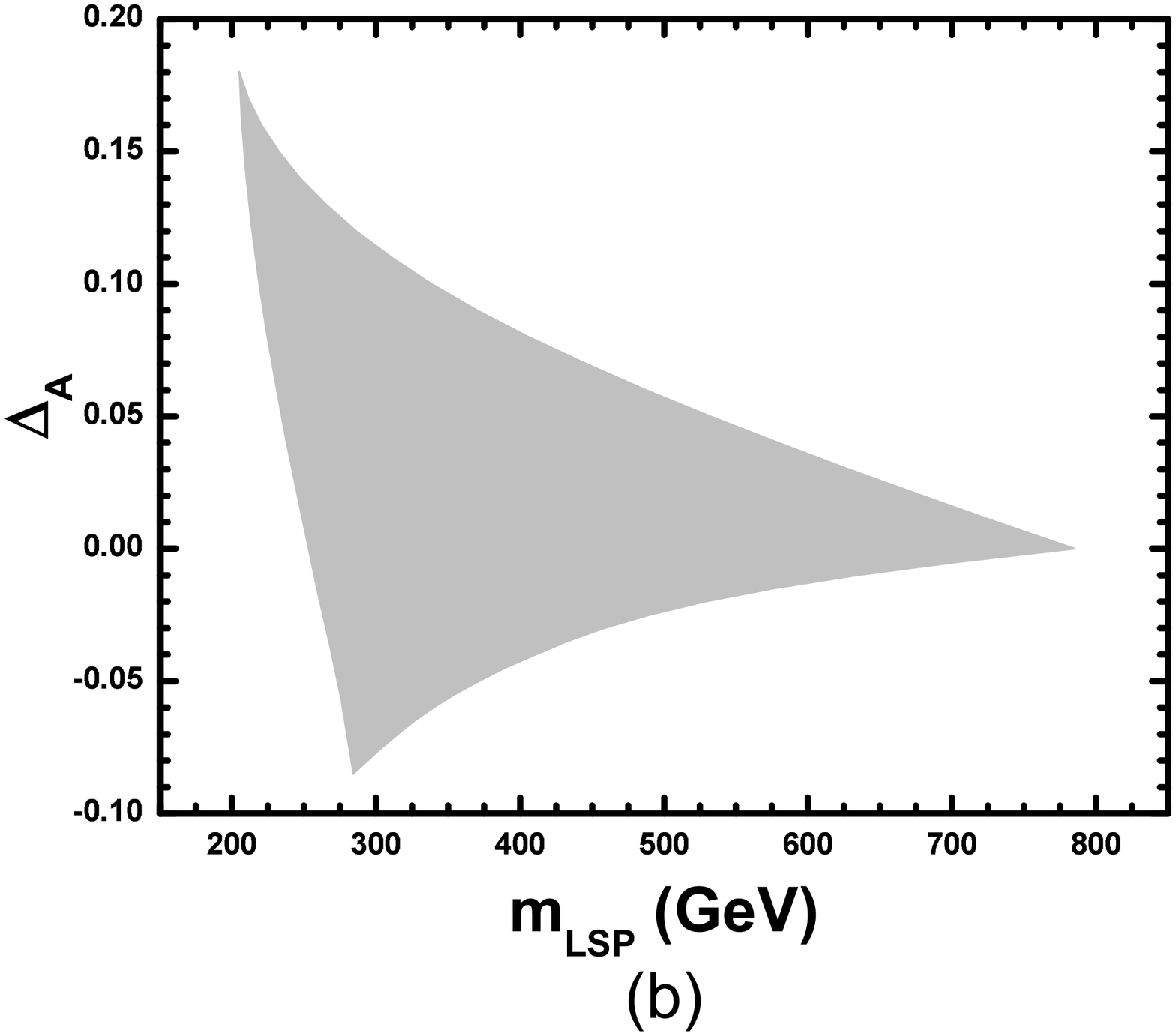,height=3.8in,angle=-90} \hfill
\end{minipage}
\hfill\hspace*{-.71in}
\begin{minipage}{8in}
\epsfig{file=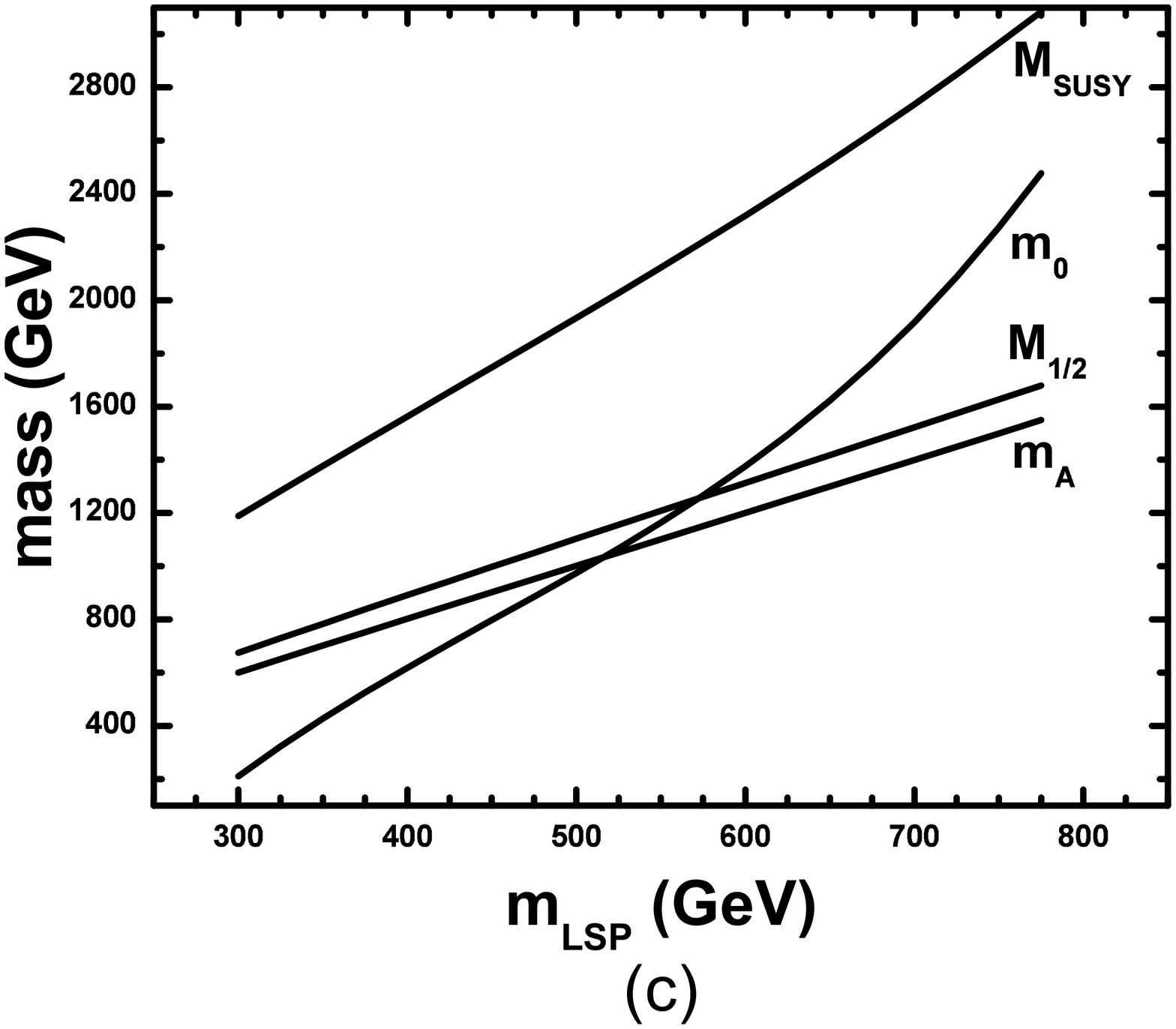,height=3.8in,angle=-90} \hspace*{-1.37 cm}
\epsfig{file=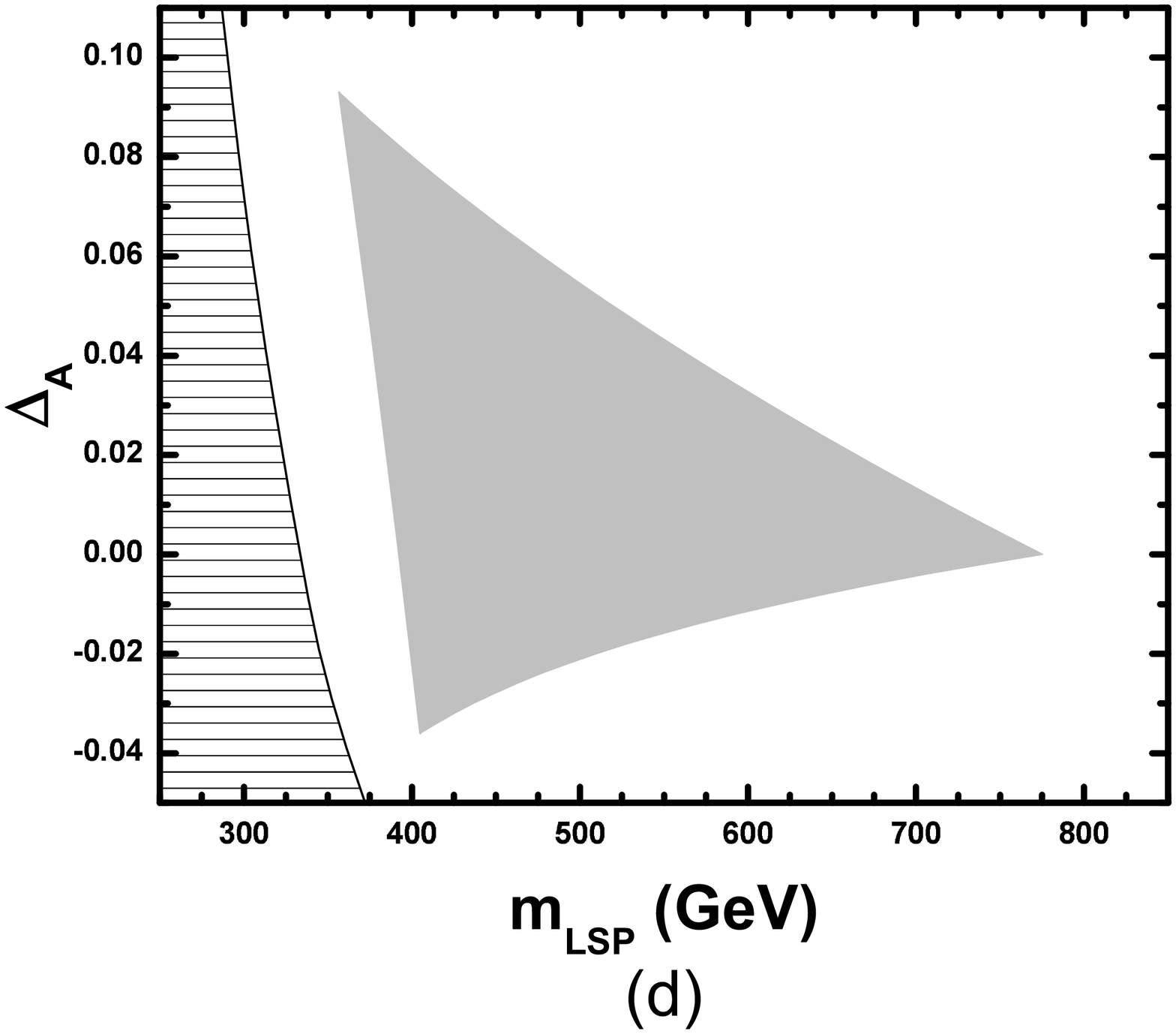,height=3.8in,angle=-90} \hfill
\end{minipage}
\hfill \caption[]{\sl For $r_{\tilde f}=r_2=r_3=1$ and high [low]
$m_0$, the mass parameters $M_{1/2}$, $m_0$, $m_A$ and $M_{\rm
SUSY}$ versus $m_{\rm LSP}$ for $\Delta_A=0$ {\sf (a [c])} and the
allowed (shaded) area on the $m_{\rm LSP}-\Delta_{A}$ plane {\sf
(b [d])}. Ruled is, also, the area favored by the optimistic upper
bound on $m_{\rm LSP}$ from Eq. (\ref{g2e}{\sf a}) }\label{5pa}
\end{figure}

\subsection{$A$-P{\ssz OLE} E{\ssz FFECT}
({\ssz WITH OR WITHOUT} $\nta-\chb-\ntb$ C{\ssz
OANNIHILATIONS})}\label{resulta}

\hspace{.562cm} For $r_2=r_3=1$ and $0<r_{\tilde f}\leq1.2$,
reduction of $\Omega_{\rm LSP}h^2$ caused by the $A$PE is
possible. Especially, for $r_{\tilde f}=1$, there are two
different combinations of $M_{1/2}$ and $m_0$, which support this
possibility. In Figs. \ref{5pa}-{\sf (a)} and \ref{5pa}-{\sf (c)},
we present the mass parameters $M_{1/2}$, $m_0$, $m_A$  and
$M_{\rm SUSY}$ versus $m_{\rm LSP}$ for $\Delta_A=0$ in these two
cases. We observe that the main difference between them is related
to the value of $m_0$, which turns out to be relatively high [low]
in Fig. \ref{5pa}-{\sf (a [c])}. In these, the various lines
terminate at low [high] $m_{\rm LSP}$'s due to improper RESB
($m_A^2<0$) [$m^{\rm c}_b(M_Z)$].

The corresponding allowed areas on the $m_{\rm LSP}-\Delta_A$
plane are displayed in Figs. \ref{5pa}-{\sf (b)} and
\ref{5pa}-{\sf (d)}. In both cases, the left (almost vertical)
boundary of the allowed (shaded) region comes from Eq.
(\ref{bsgb}{\sf b}) while the lower and upper curved boundaries
correspond to the saturation of Eq. (\ref{cdmb}). A simultaneous
satisfaction of Eq. (\ref{g2e}{\sf a}) is impossible. More
explicitly, we find the following allowed ranges:
\beq {\bf i.}~~ 205~[284]~{\rm GeV}\lesssim m_{\rm
LSP}\lesssim784~{\rm GeV}~~\mbox{for}~~ 0.18~\left[|-0.085|\right]
\gtrsim\left|\Delta_A\right|\gtrsim0,\label{5Au}\eeq
with $45~[45.2]\gtrsim\tan\beta \gtrsim41.2$ in the high $m_0$
case (Fig. \ref{5pa}-{\sf (b)}). Saturation of the optimistic
bound from Eq. (\ref{g2e}{\sf a}) is not possible in the overall
investigated parameter space.
\beq {\bf ii.} ~~357~[405]~{\rm GeV}\lesssim m_{\rm
LSP}\lesssim775~{\rm GeV}~~\mbox{for}~~
0.093~[|-0.036|]\gtrsim\left|\Delta_A\right|\gtrsim0
\label{5Ad}\eeq
with $36~[35.2]\lesssim \tan\beta\lesssim 39.8$ in the low $m_0$
case (Fig. \ref{5pa}-{\sf (d)}). The bound of Eq. (\ref{g2e}{\sf
a}) implies $m_{\rm LSP}\lesssim372~{\rm GeV}$.

\begin{figure}[t]
\hspace*{-.71in}
\begin{minipage}{8in}
\epsfig{file=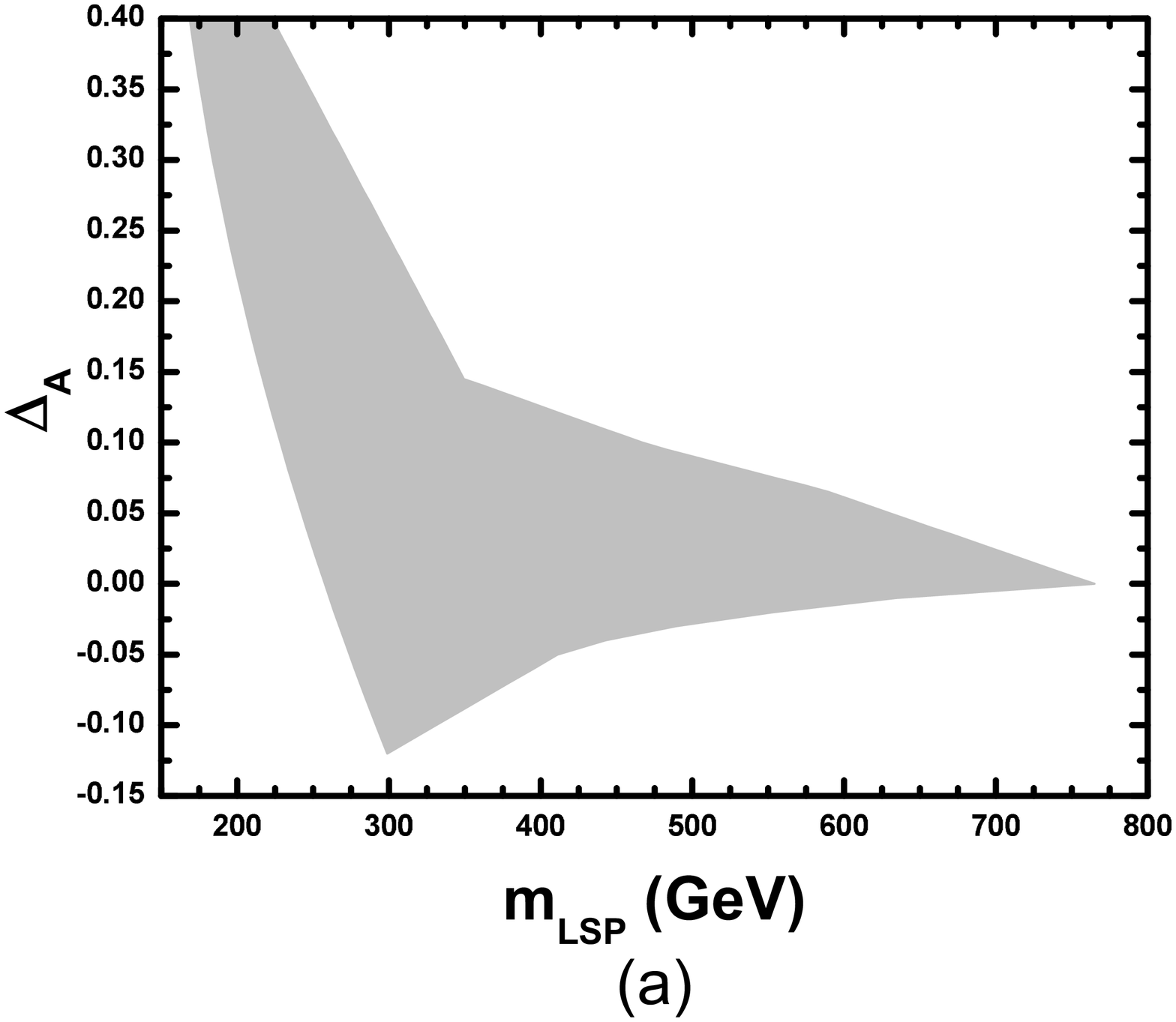,height=3.8in,angle=-90} \hspace*{-1.37 cm}
\epsfig{file=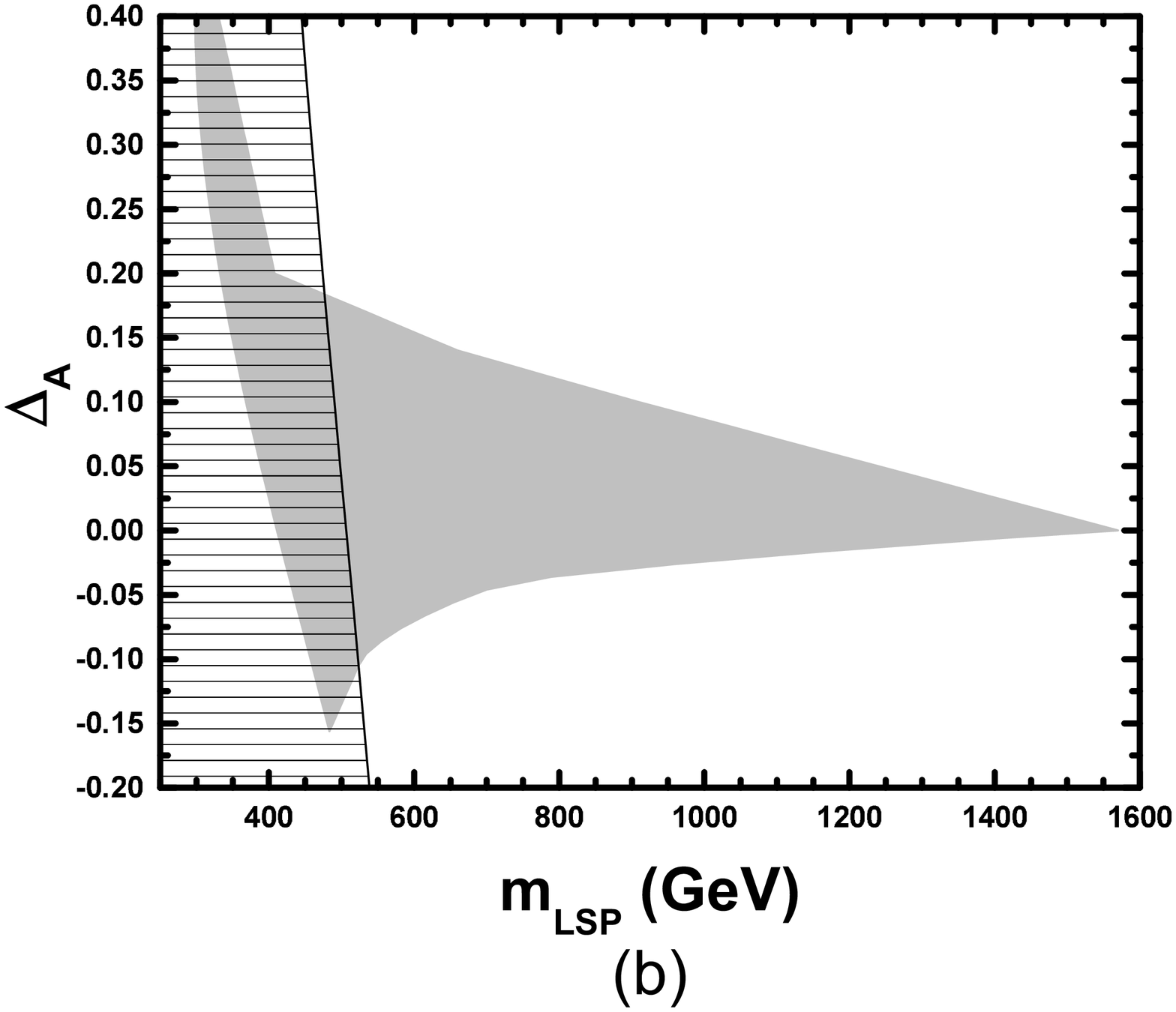,height=3.8in,angle=-90}
\end{minipage}
\hfill \caption[]{\sl The allowed (shaded) area on the $m_{\rm
LSP}-\Delta_{A}$ plane for $r_{\tilde f}=1, r_2=0.6, r_3=1$ and
high $m_0$ {\sf (a)}, $r_{\tilde f}=0.4, r_2=r_3=0.6$ and low
$m_0$ {\sf (b)}. Ruled is, also, the area favored by the
optimistic upper bound on $m_{\rm LSP}$ from Eq. (\ref{g2e}{\sf
a}).}\label{5pb}
\end{figure}

Comparing Figs. \ref{5pa}-{\sf (b)} and \ref{5pa}-{\sf (d)}, we
observe that the allowed area for low $m_0$ can be included in
this for high $m_0$, with the lower and upper curved boundaries
being almost identical. The main difference is that the bound from
Eq. (\ref{bsgb}{\sf b}) is more restrictive in the low $m_0$ case,
due to the lighter stop spectrum. This difference is also shown in
Fig. \ref{csec}-{\sf (a)}, where we depict $\sigma_{\tilde\chi
p}^{\rm SI}$ versus $m_{\rm LSP}$ for $\Delta_A=0.1$ and low
[high] $m_0$ (grey [light grey]) band. Obviously the high $m_0$ is
phenomenologically more attractive.

Coexistence of $A$PE and $\nta-\chb-\ntb$ CAM can further enlarge
the allowed areas. Indeed, analyzing two cases, we find
\beq {\bf i^\prime.}~~ 299~{\rm GeV}\lesssim m_{\rm
LSP}\lesssim765~{\rm GeV}~~\mbox{for}~~ |-0.12|
\gtrsim\left|\Delta_A\right|\gtrsim0,\label{5Aun}\eeq
with $35.5\lesssim\tan\beta\lesssim37.2$ and high $m_0$ for
$r_{\tilde f}=1, r_2=0.6$ and $r_3=1$ (Fig. \ref{5pb}-{\sf (a)})
without possibility of saturation of the optimistic bound from Eq.
(\ref{g2e}{\sf a}). The obtained $\sigma_{\tilde\chi p}^{\rm SI}$
turns out to be quite similar to this of light grey band in Fig.
\ref{csec}-{\sf (a)}.
\beq {\bf ii^\prime.} ~~483.5~{\rm GeV}\lesssim m_{\rm
LSP}\lesssim1570~{\rm GeV}~~\mbox{for}~~
|-0.156|]\gtrsim\left|\Delta_A\right|\gtrsim0 \label{5Adn}\eeq
with $35\lesssim\tan\beta\lesssim39.4$ and low $m_0$ for
$r_{\tilde f}=0.4$ and $r_2=r_3=0.6$ (Fig. \ref{5pb}-{\sf (b)}).
The bound of Eq. (\ref{g2e}{\sf a}) can be satisfied for $m_{\rm
LSP}\lesssim538~{\rm GeV}$ due to low $r_{\tilde f}$. The
corresponding $\sigma_{\tilde\chi p}^{\rm SI}$ is also increased
due to stronger higgsino component of the LSP as is shown in Fig.
\ref{csec}-{\sf (a)}, cyan band.

Due to the existence of the $\nta-\chb-\ntb$ CAM, in both latter
(i$^\prime$, ii$^\prime$) cases, there is no upper [lower] bound
on $\Delta_A~[m_{\rm LSP}]$, for $\Delta_A>0$, contrary to the
former cases (i, ii). Evident is, also, in any case that the
$\Omega_{\rm LSP}h^2$ reduction, because of the $A$PE, is more
efficient for $\Delta_A>0$ than for $\Delta_A<0$, in accord with
the findings of Ref. \cite{nra}.

\begin{figure}[t]
\hspace*{-.71in}
\begin{minipage}{8in}
\epsfig{file=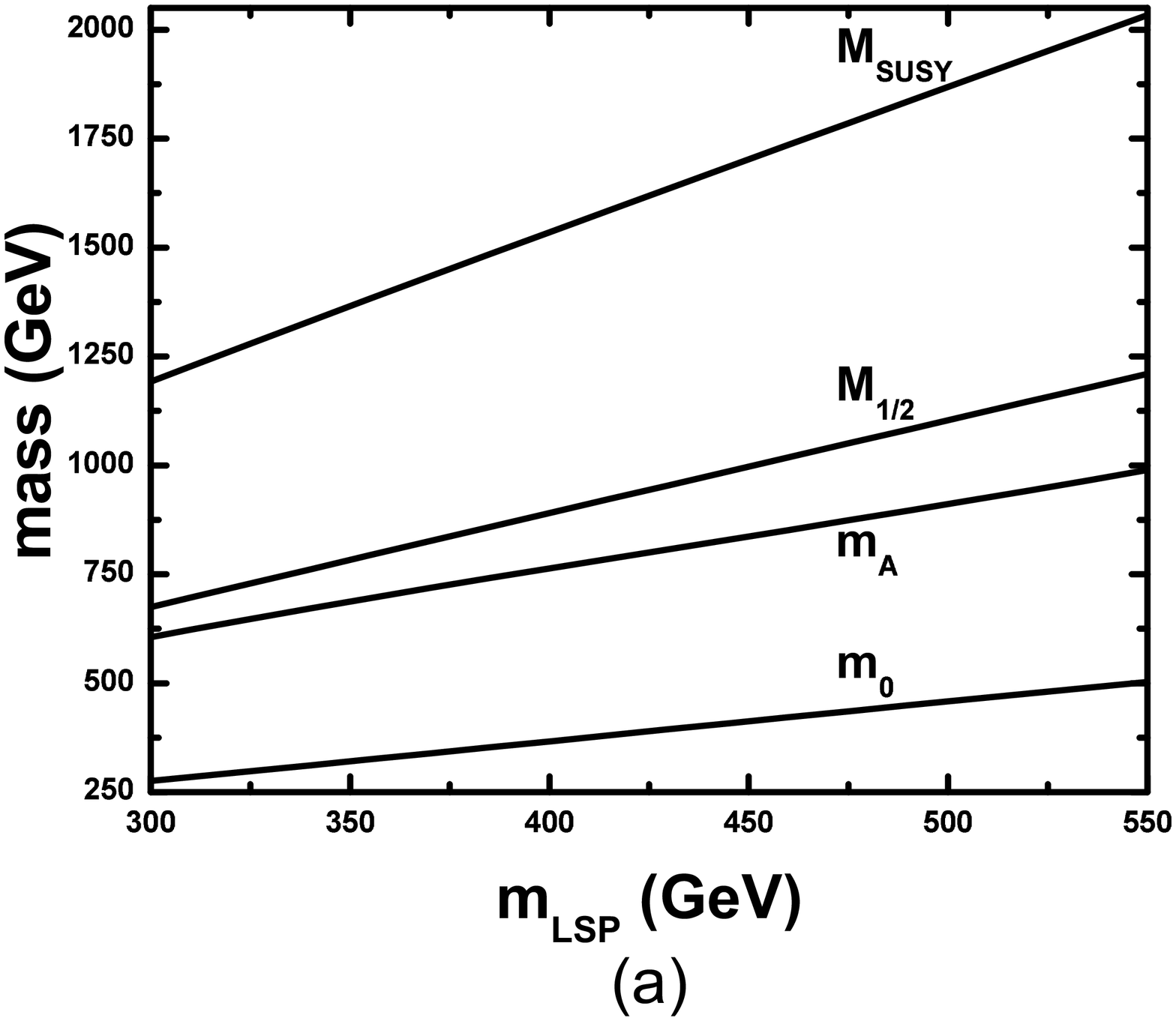,height=3.8in,angle=-90} \hspace*{-1.37 cm}
\epsfig{file=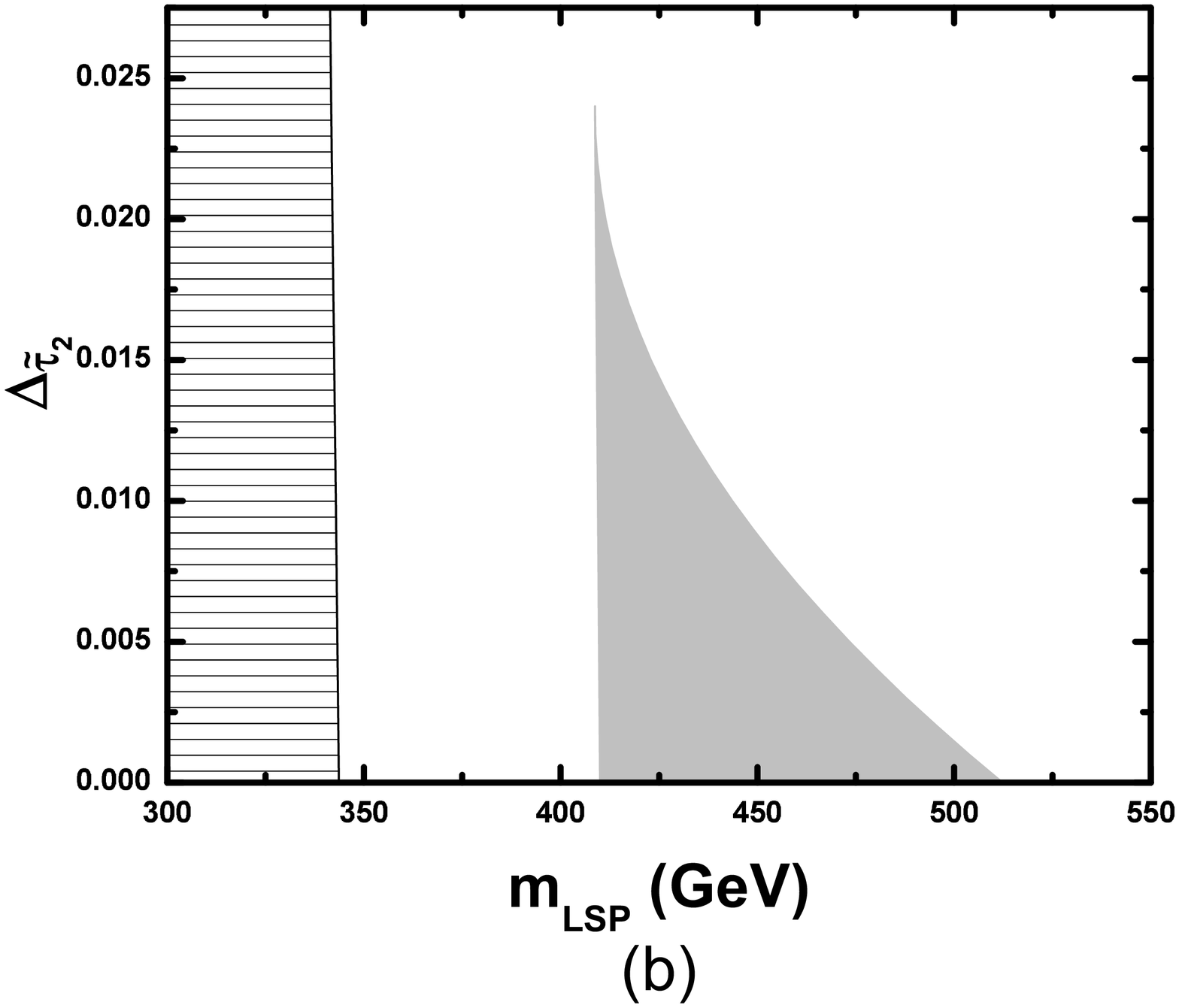,height=3.8in,angle=-90} \hfill
\end{minipage}
\hfill\hspace*{-.71in}
\begin{minipage}{8in}
\epsfig{file=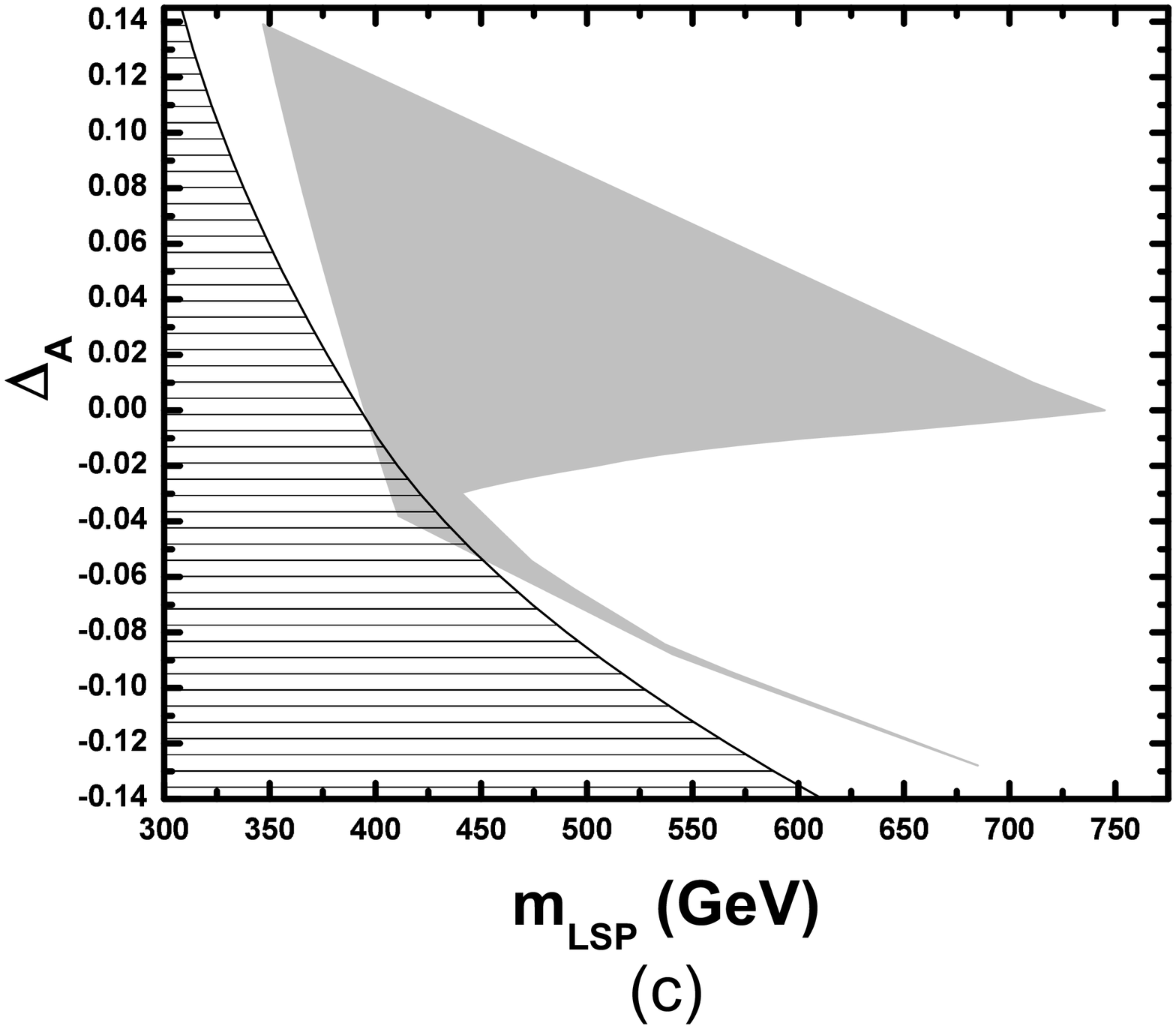,height=3.8in,angle=-90} \hspace*{-1.37 cm}
\epsfig{file=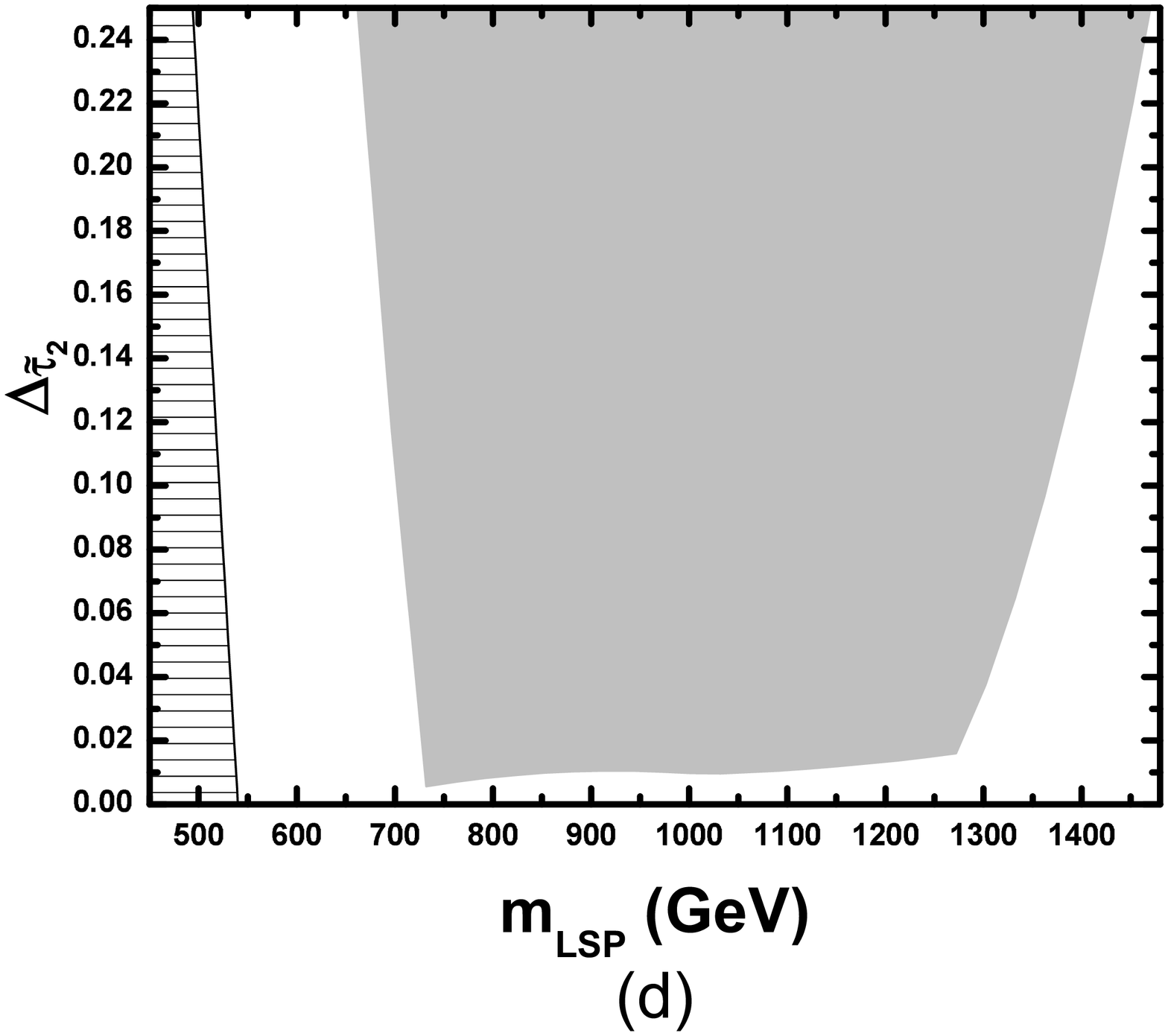,height=3.8in,angle=-90} \hfill
\end{minipage}
\hfill \caption[]{\sl The mass parameters $M_{1/2}$, $m_0$, $m_A$
and $M_{\rm SUSY}$ versus $m_{\rm LSP}$ for $r_{\tilde
f}=r_2=r_3=1$ and $\Delta_{\tilde\tau_2}=0$ {\sf (a)} and the
allowed (shaded) area for $r_{\tilde f}=1$ and $r_2=1~[0.6],
r_3=1~[0.5]$ {\sf (b [d])} on the $m_{\rm
LSP}-\Delta_{\tilde\tau_2}$ plane or $r_{\tilde f}=1$ and
$r_2=0.6, r_3=1$ on the $m_{\rm LSP}-\Delta_{A}$ plane {\sf (c)}.
Ruled is, also, the area favored by the optimistic upper bound on
$m_{\rm LSP}$ from Eq. (\ref{g2e}{\sf a}).}\label{5st}
\end{figure}

In both cases, the relative contributions beyond 5$\%$ of the
(co)annihilation processes to $\Omega_{\rm LSP}h^2$ as $m_{\rm
LSP}$ and $\Delta_A$ vary in the ranges of Eq. (\ref{5Au}) or
(\ref{5Ad}) [(\ref{5Aun}) or (\ref{5Adn})], are:
\begin{center}
\begin{tabular}{ll}
$\nta\nta\rightarrow b\bar b$ &$(87-86)~[52-78]\%$\\
$\nta\nta\rightarrow \tau\bar\tau $&$(12.5-14)~[7-13]\%$
\end{tabular}
\end{center}


If we had imposed UGMs with $\mu<0$ and $r_{\tilde f}=1$, the high
$m_0$ case would not have survived, due to larger $\tan\beta$
which would have invalidated the RESB, while for low $m_0$, we
could have found allowed area similar to this in Fig.
\ref{5pb}-{\sf (b)} with $m_{\rm LSP}\gtrsim466~{\rm GeV}$ and
$\tan\beta\simeq42$.


\subsection{$\nta(-\chb-\ntb)-\stau$ C{\ssz OANNIHILATIONS}}
\label{resultt}

\hspace{.562cm} The most usual and easily achieved (for $r_{\tilde
f}\neq1$, also) CAM is the $\nta-\stau$ CAM. The mass parameters
$M_{1/2}$, $m_0$, $m_A$ and $M_{\rm SUSY}$ which support this
situation, are plotted versus $m_{\rm LSP}$ for
$\Delta_{\tilde\tau_2}=0$ and $r_{\tilde f}=r_2=r_3=1$ in Fig.
\ref{5st}-{\sf (a)}. We observe that $M_{1/2} \gg m_0$, unlike all
the other cases.

The corresponding allowed area on the $m_{\rm
LSP}-\Delta_{\tilde\tau_2}$ plane, for the same $r$ 's is depicted
in Fig. \ref{5st}-{\sf (b)} and it turns out to be disconnected
from these of subsec. \ref{resulta}. The left (almost vertical)
[right curved] boundary of the allowed (shaded) region is derived
from Eq. (\ref{bsgb}{\sf b}~ [\ref{cdmb}]). It is obvious that
strong degeneracy among LSP and NLSP is needed in order the
criteria of Eqs. (\ref{cdmb}) and (\ref{bsgb}{\sf b}) to be
simultaneously fulfilled without a possible achievement of Eq.
(\ref{g2e}{\sf a}), since it implies $m_{\rm LSP}\lesssim343~{\rm
GeV}$ (Table 1, left column). Consequently, the corresponding
$\sigma_{\tilde\chi p}^{\rm SI}$ lies also well below the range of
Eq. (\ref{sgmb}{\sf b}), as shown in Fig. \ref{csec}-{\sf (b)},
dark grey band.

\begin{table}[!t]
\caption {\bf D\ssz OMINANT \nsz C\ssz ONTRIBUTIONS TO \nsz
{\boldmath $\Omega_{\rm LSP}h^2$} }
\begin{center}
\begin{tabular}{|l||c|c|c|}\hline
\multicolumn{4}{|c|}{\bf M\ssz ODEL \nsz P\ssz ARAMETERS} \\
\hline
{$r_{\tilde f}$}&{$1$}&{$1$}&{$1$}\\
{$r_2, r_3$}&{$r_2=r_3=1$}&{$r_2=0.6, r_3=1$}&{$r_2=0.6,
r_3=0.5$}\\ \hline \hline
\multicolumn{4}{|c|}{\bf A\ssz LLOWED \nsz R\ssz ANGES}\\ \hline
$\tan\beta$&{$35$}&{${\it 34.1}-35.1$}&{$34.2-{\it 39.4}$} \\
$m_{\rm LSP}~({\rm GeV})$ &{$408-512$} &{${\it
400}-686$}&{$734-{\it 1420}$}\\
$\Delta_{\tilde \tau_2}$&{$0.15-0$}&{${\it 0.25}-0$}&{$0.01-{\it
0.25}$}\\ \hline \hline
\multicolumn{4}{|c|}{\bf P\ssz ROCESSES \nsz W\ssz HICH \nsz C\ssz
ONTRIBUTE \nsz M\ssz ORE \nsz T\ssz HAN \nsz {\boldmath $7\%$}}
\\ \hline \multicolumn{1}{|l||}{\bf P\ssz ROCESS} &
\multicolumn{3}{|c|}{\bf C\ssz ONTRIBUTION \nsz ({\boldmath
$\%$})}\\ \cline{2-4} \hline
$\nta\nta\rightarrow b\bar b$ &$40-2$&$82-0$&$-$\\
$\nta\nta\rightarrow \tau\bar \tau$ &$8-1$&$13-0$&$-$\\
$\nta\nta\rightarrow W^+W^-$ &$-$&$-$&$7-0$\\
$\tilde\chi\tilde\tau_2\rightarrow\tau\gamma$ &$11-13$&$-$&$-$\\
\hline
$\tilde\tau_2\tilde\tau_2\rightarrow\tau\tau$&$10.5-32$&$0-12$&$-$\\
$\tilde\tau_2\tilde\tau_2^\ast\rightarrow\gamma\gamma$&$3-9.6$&$-$&$-$\\
$\tilde\tau_2\tilde\tau_2^\ast\rightarrow\gamma
Z$&$1.6-7$&$-$&$-$\\
$\tilde\tau_2\tilde\tau_2^\ast\rightarrow b\bar b$
&$12-18$&$0-18$&$-$\\
$\tilde\tau_2\tilde\tau_2^\ast\rightarrow t\bar t$
&$-$&$0-16$&$-$\\ \hline
$\nta\chbp\rightarrow f_u \bar f_d $&$-$&$-$&$16.5-27$\\
$\chbp\chbm\rightarrow f_u \bar f_d $&$-$&$-$&$0-12$\\
$\ntb\chbp\rightarrow f_u \bar f_d $&$-$&$-$&$0-15$
%
\\ \hline
\end{tabular}
\end{center}
\end{table}
These ``pessimistic'' results are not essentially alleviated,
lifting the UGMs. Indeed, for $r_{\tilde f}=1, r_2=0.6$ and
$r_3=1$ the allowed area is somehow enlarged to higher $m_{\rm
LSP}$ due to extra CAMs, but the lower $m_{\rm LSP}$ is still high
enough to be phenomenologically interesting (Table 1, middle
column, italic numbers are referred to indicative and not absolute
bounds). In this case the $\nta-\stau$ coannihilation tail turns
out to be connected to the allowed region caused by $A$PE. So, for
economy, they are collectively presented on the $m_{\rm
LSP}-\Delta_{A}$ plane in Fig. \ref{5st}-{\sf (c)}. On the other
hand, for $r_{\tilde f}=1,~r_2=0.6$ and $r_3=0.5$, an unusual
behavior on the $m_{\rm LSP}-\Delta_{\tilde\tau_2}$ plane is
presented in Fig. \ref{5st}-{\sf (d)}. There, $\nta-\chb-\ntb$ CAM
are more efficient than the $\nta-\stau$. Strengthening the
$\nta-\stau$ proximity, the $\nta-\chb-\ntb$ CAM contribution to
$\Omega_{\rm LSP}h^2$ decreases (Table 1, right column). So,
$\Omega_{\rm LSP}h^2$ increases due to the domination of the
weaker $\nta-\stau$ CAM and a lower bound on the $m_{\rm
LSP}-\Delta_{\tilde\tau_2}$ plane emerges.

It is worth mentioning, that had we assumed UGMs with $\mu<0$ and
$r_{\tilde f}=1$, the lower bound on $m_{\rm LSP}$, derived again
from Eq. (\ref{bsgb}{\sf b}), would have been much more
restrictive ($m_{\rm LSP}\gtrsim519~{\rm GeV}$ for
$\tan\beta\simeq40.6$). An explanation is given in the
introduction of sec. \ref{result}. Thus, we would have been
practically left without allowed area.

\begin{figure}[t]
\hspace*{-.71in}
\begin{minipage}{8in}
\epsfig{file=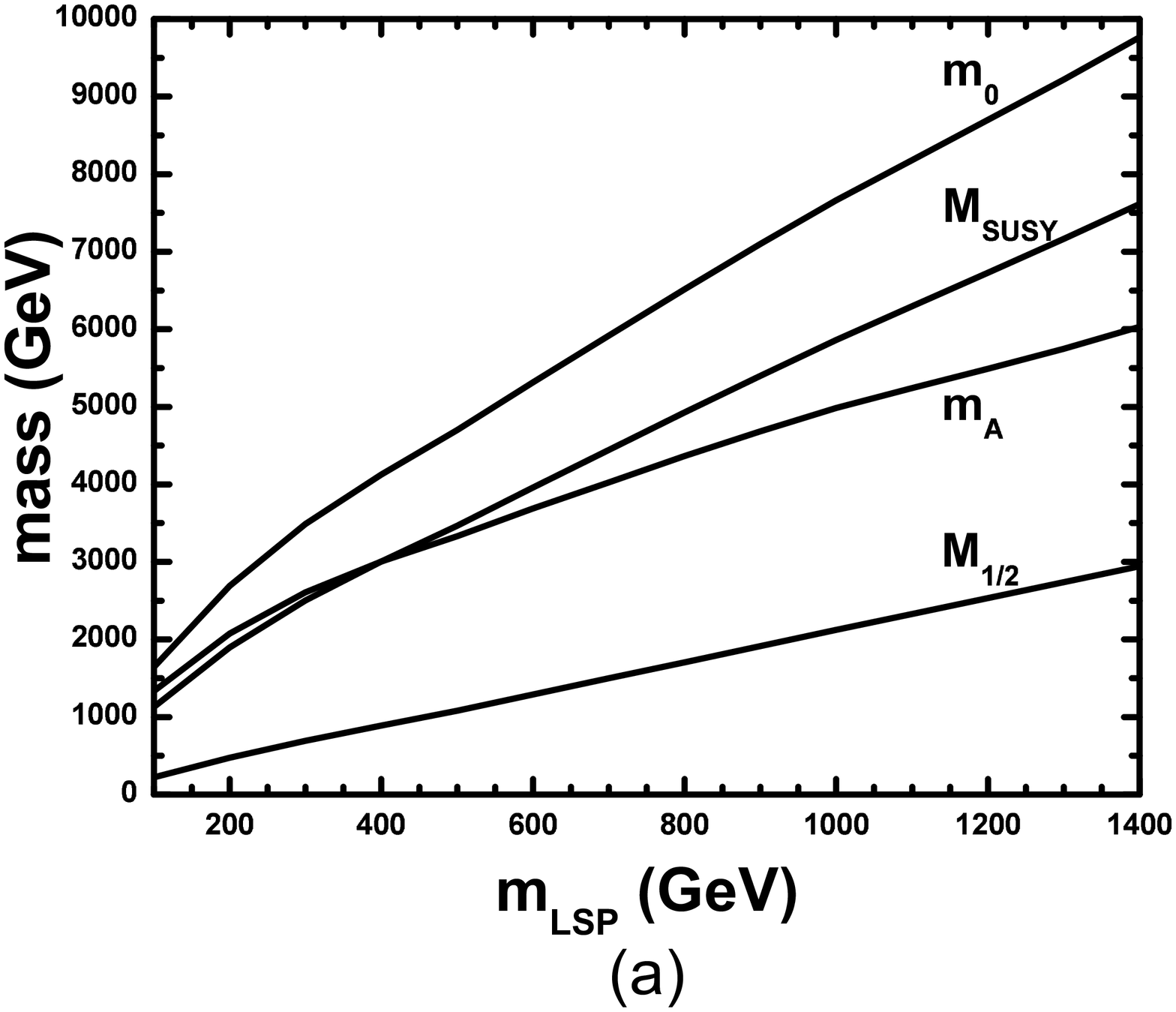,height=3.8in,angle=-90} \hspace*{-1.37 cm}
\epsfig{file=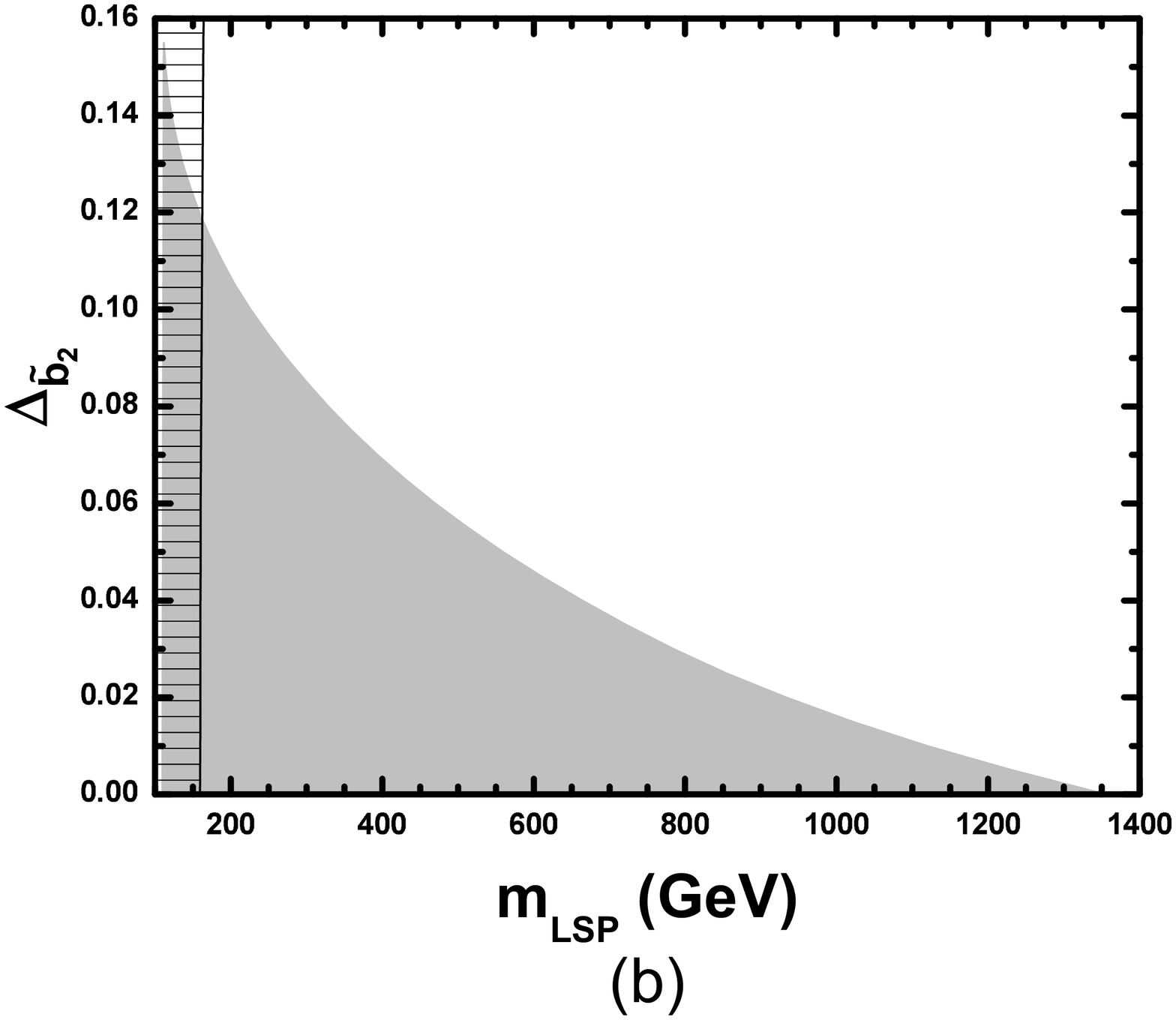,height=3.8in,angle=-90} \hfill
\end{minipage}
\hfill\hspace*{-.71in}
\begin{minipage}{8in}
\epsfig{file=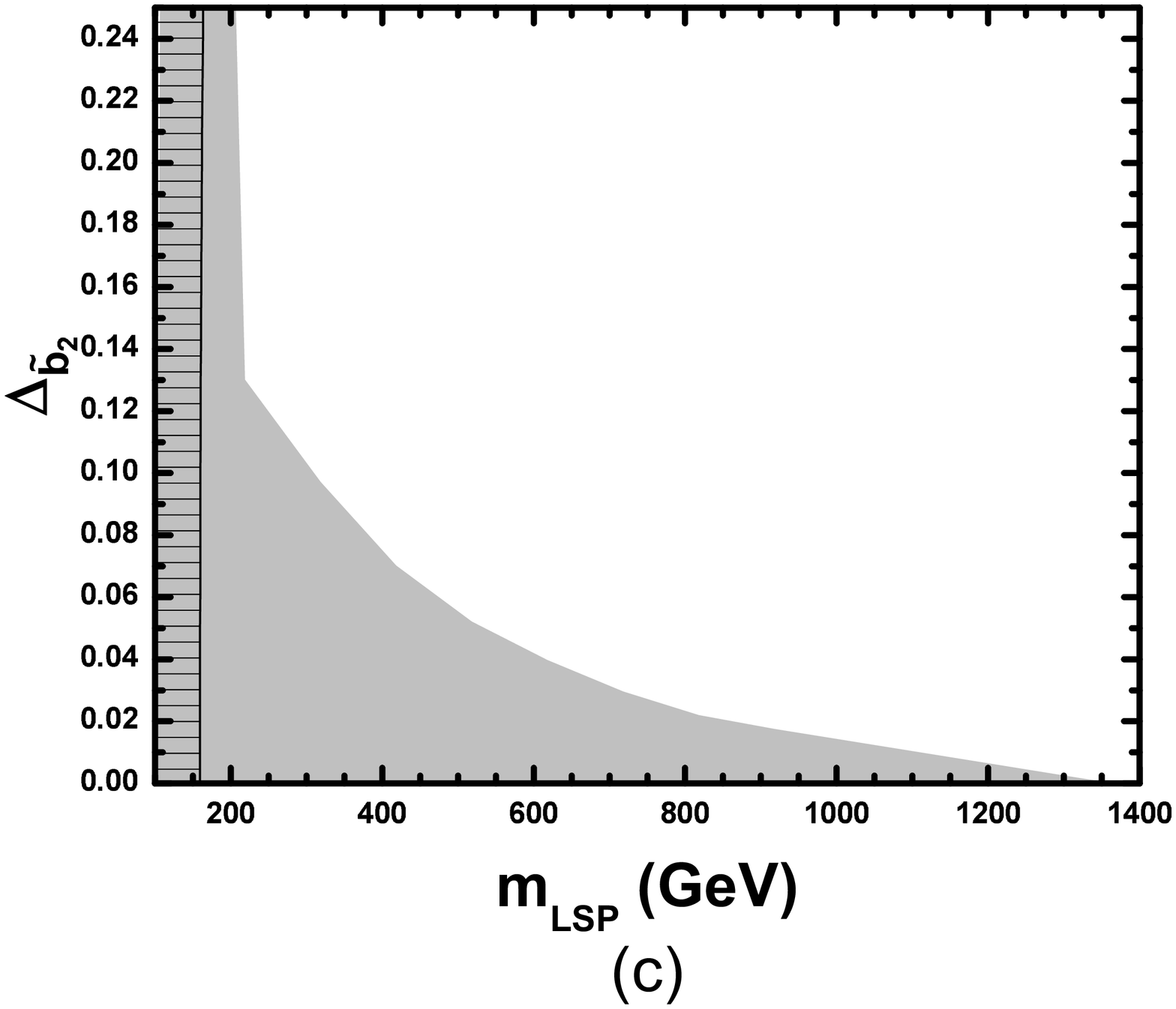,height=3.8in,angle=-90} \hspace*{-1.37 cm}
\epsfig{file=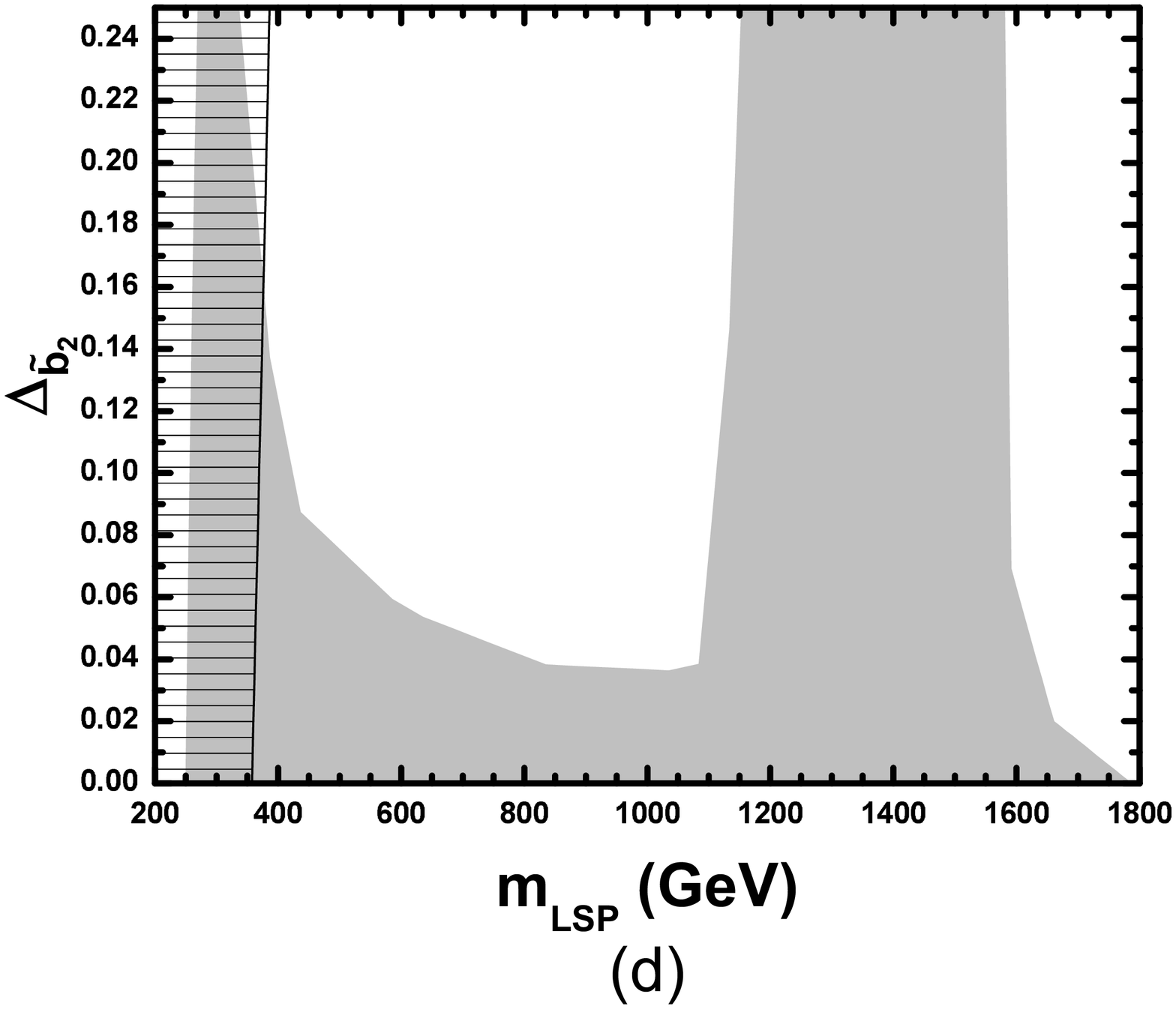,height=3.8in,angle=-90} \hfill
\end{minipage}
\hfill \caption[]{\sl The mass parameters $M_{1/2}$, $m_0$, $m_A$
and $M_{\rm SUSY}$ versus $m_{\rm LSP}$ for $r_{\tilde f}=0.4,
r_2=r_3=1$ and $\Delta_{\tilde b_2}=0$ {\sf (a)} and the allowed
(shaded) area on the $m_{\rm LSP}-\Delta_{\tilde b_2}$ plane for
$r_{\tilde f}=0.4$ and $r_2=r_3=1$ {\sf (b)}, $r_2=0.6, r_3=1$
{\sf (c)}, $r_2=0.6, r_3=0.5$ {\sf (d)}. Ruled is, also, the area
favored by the optimistic upper bound on $m_{\rm LSP}$ from Eq.
(\ref{g2e}{\sf a}).}\label{5sb}
\end{figure}


\subsection{$\nta(-\chb-\ntb)-\sbb$ C{\ssz
OANNIHILATIONS}}\label{resultb}

\hspace{.562cm} For moderate to low $r_{\tilde f}$, a new type of
CAM between \nta\ and \sbb\ can be obtained. However, the needed
mass proximity can be established only for $M_{1/2}\ll m_0$, with
$m_0>1 ~{\rm TeV}$ (in accord with the findings of Ref.
\cite{santoso}). This is, because $m_{\tilde Q}$ and $m_{\tilde
b^c}$ [$m_{\tilde L}$ and $m_{\tilde \tau^c}$] de[in]-crease as
$m_0$ increases, since $h_t$ and $h_b$ [$h_\tau$] in[de]-crease
with the running from $M_{\rm GUT}$ to $M_Z$ \cite{rge}.
Therefore, $m_{\tilde b_2}~[m_{\tilde \tau_2}]$ de[in]-creases
drastically. So, $\tilde b_2$, which is mainly $\tilde b^c$
(unlike the similar case of Ref. \cite{santoso}), can become
coannihilator of \nta.

A typical example for the values of $M_{1/2}$, $m_0$, $m_A$  and
$M_{\rm SUSY}$ versus $m_{\rm LSP}$ in this case, is presented in
Fig. \ref{5sb}-{\sf (a)} for $\Delta_{\tilde b_2}=0$ and
$r_{\tilde f}=0.4$ and $r_2=r_3=1$. The resulting allowed area on
the $m_{\rm LSP}-\Delta_{\tilde b_2}$ plane for the same $r$'s is
displayed in Fig. \ref{5sb}-{\sf (b)}. The left (almost vertical)
[right curved] boundary of the allowed (shaded) region is derived
from Eq. (\ref{mhb}{\sf a}~[\ref{cdmb}]). The bound of the right
curve can be occasionally relaxed for $r_2<1$ and/or $r_3<1$. Two
examples are depicted in Fig. \ref{5sb}-{\sf (c~[d])} for
$r_{\tilde f}=0.4$ and $r_2 =0.6, r_3=1~[0.5]$. Due to extra CAMs
there are regions without upper bound on $\Delta_{\tilde b_2}$.
However the left (almost vertical) boundary becomes gradually more
restrictive derived, in these cases, from Eq. (\ref{bsgb}{\sf b}).
Our findings are in detail and comparatively listed in Table 2
(recall that the italic numbers are refereed to indicative
bounds). Optimistic bounds on $m_{\rm LSP}$ derived from Eq.
(\ref{g2e}{\sf a}) are included in parenthesis.

It should be emphasized that the reduction of $\Omega_{\rm
LSP}h^2$ caused by $\nta-\sbb$ CAM is much more efficient than
this by $\nta-\stau$ and $\nta-\ntn-\stau$ CAM. As a consequence,
larger $m_{\rm LSP}$'s (and $\Delta_{\tilde b_2}$'s) are allowed.
It is also always constructive to and stronger than a possible
$\nta-\chb-\ntb$ CAM (for the used $r$ 's) contrary to the case of
$\nta-\chb-\ntb-\stau$ CAM (sec. \ref{result}). At the same time,
thanks to heavier stop and higgs sector, the satisfaction of Eq.
(\ref{bsgb}{\sf b}) is facilitated and there is parameter space
where the putative bound of Eq. (\ref{g2e}{\sf a}) can be
fulfilled, too.

\begin{table}[!t]
\caption {\bf D\ssz OMINANT \nsz C\ssz ONTRIBUTIONS TO \nsz
{\boldmath $\Omega_{\rm LSP}h^2$} }
\begin{center}
\begin{tabular}{|l||c|c|c|}\hline
\multicolumn{4}{|c|}{\bf M\ssz ODEL \nsz P\ssz ARAMETERS} \\
\hline
{$r_{\tilde f}$}&{$0.4$}&{$0.4$}&{$0.4$}\\
{$r_2, r_3$}&{$r_2=r_3=1$}&{$r_2=0.6, r_3=1$}&{$r_2=0.6,
r_3=0.5$}\\ \hline \hline
\multicolumn{4}{|c|}{\bf A\ssz LLOWED \nsz R\ssz ANGES}\\ \hline
$\tan\beta$&{$33.5-41.7$}&{$33.5-41.2$}&{$34.2-41.4$} \\
$m_{\rm LSP}~({\rm GeV})$ &{$111-(160)1354$}
&{$104-(164)1357$}&{$251-(386)1785$}\\
$\Delta_{\tilde b_2}$&{$0.15-0$}&{${\it 0.25}-0$}&{${\it
0.25}-0$}\\ \hline \hline
\multicolumn{4}{|c|}{\bf P\ssz ROCESSES \nsz W\ssz HICH \nsz C\ssz
ONTRIBUTE \nsz M\ssz ORE \nsz T\ssz HAN \nsz {\boldmath $5\%$}}
\\ \hline \multicolumn{1}{|l||}{\bf P\ssz ROCESS} &
\multicolumn{3}{|c|}{\bf C\ssz ONTRIBUTION \nsz ({\boldmath
$\%$})}\\ \cline{2-4} \hline
$\nta\nta\rightarrow t\bar t$ &$38.5-0$&$-$&$10-0$\\
$\nta\nta\rightarrow b\bar b$ &$37.6-0$&$9-0$&$19-0$\\
$\nta\nta\rightarrow W^-W^+$ &$38.5-0$&$8-0$&$15-0$\\ \hline
$\nta\sbb\rightarrow g b $&$37.3-5.6$&$-$&$0-9.5$\\
$\nta\sbb\rightarrow W^- t $&$-$&$-$&$0-7$\\ \hline
$\sbb\sbb\rightarrow bb $&$1.5-7.1$&$0-8$&$0-5$\\
$\sbb\sbbs\rightarrow gg $&$19-79$&$0-79$&$0-19$\\ \hline
$\nta\chbp\rightarrow f_u \bar f_d $&$-$&$27-0$&$19-7$\\
$\chbp\chbm\rightarrow f \bar f $&$-$&$14-0$&$6-4$\\
$\nta\ntb\rightarrow W^-W^+ $&$-$&$5-0$&$5-0$\\
$\ntb\chbp\rightarrow f_u \bar f_d $&$-$&$17-0$&$7-0$
\\ \hline
\end{tabular}
\end{center}
\end{table}

\begin{figure}[t]
\hspace*{-.71in}
\begin{minipage}{8in}
\epsfig{file=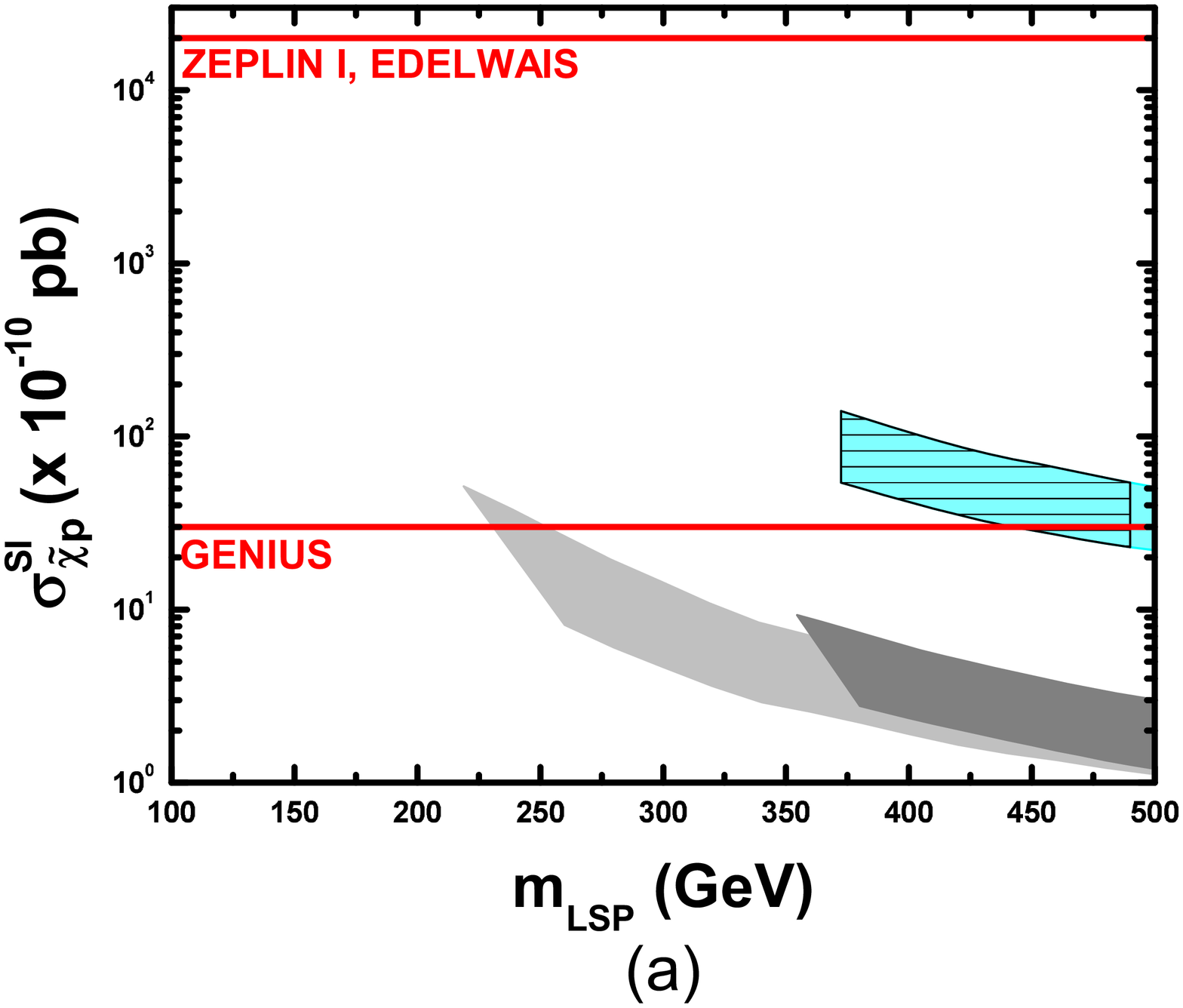,height=3.8in,angle=-90} \hspace*{-1.37
cm} \epsfig{file=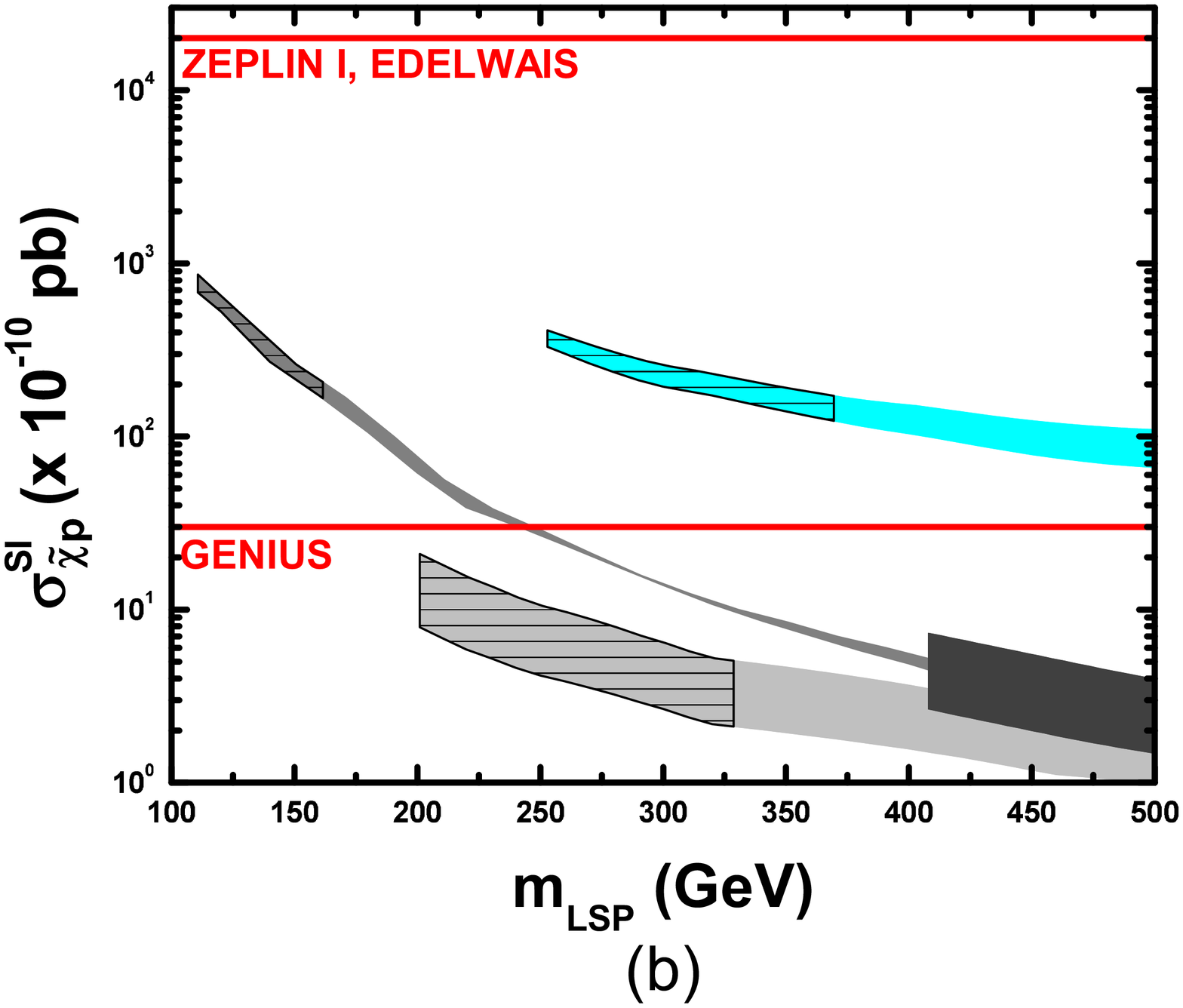,height=3.8in,angle=-90} \hfill
\end{minipage}
\hfill \caption[]{\sl $\sigma_{\tilde\chi p}^{\rm SI}$ versus
$m_{\rm LSP}$ for {\sf (a)} $\Delta_{A}=0.1$ and high (light grey
band, $r_{\tilde f}=1$) or low (grey, $r_{\tilde f}=1$, and cyan,
$r_{\tilde f}=0.4$, bands) $m_0$ {\sf (b)} $\Delta_{\tilde
b_2}=0.1$ for $r_{\tilde f}=0.4$ (gray and cyan bands),
$\Delta_{\tilde\nu_ \tau}=0$ for $r_{\tilde f}=0.2$ (light gray
band) and $\Delta_{\tilde \tau_2}=0$ for $r_{\tilde f}=1$ (dark
gray band). Gray (light, normal and dark) shaded bands are for
$r_2=r_3=1$, while cyan shaded bands are for $r_2=0.6,
r_3=0.6~[0.5]$ {\sf (a [b])}. Ruled are the areas favored by the
optimistic upper bound on $m_{\rm LSP}$ from Eq. (\ref{g2e}{\sf
a}), whereas the preferred region of $\sigma_{\tilde\chi p}^{\rm
SI}$ from various projects is, also, approximately
limited.}\label{csec}
\end{figure}

Also and more interestingly, the light lowest $m_{\rm LSP}$ has
beneficial consequences to the $\sigma_{\tilde\chi p}^{\rm SI}$
calculation. Indeed, as we can observe in Fig. \ref{csec}-{\sf
(b)}, $\sigma_{\tilde\chi p}^{\rm SI}$ for $\Delta_{\tilde
b_2}=0.1$ and even for a pure bino LSP ($r_2=r_3=1$) is enhanced
(grey band). It is almost $10^{-7}~{\rm pb}$ and lies well within
the range of Eq. (\ref{sgmb}). This is due to the increase of the
contributions, for $q=b$ (using again the notation of Ref.
\cite{drees1}): (i) $f^{(H)}_q$ (Eq. 43 of Ref. \cite{drees1})
because of the light $m_{\rm LSP}$ (ii) $B_D$ and $B_{1D}$ for
$i=2$ (Eq. 41 of Ref. \cite{drees1}), because of the $\nta-\sbb$
mass proximity. The situation remains almost unaltered for
$r_2=0.6$ and $r_3=1$ and is relatively ameliorated for $r_2=0.6$
and $r_3=0.5$, due to the sizable higgsino component of LSP. The
corresponding to the latter case band on the $m_{\rm
LSP}-\sigma_{\tilde\chi p}^{\rm SI}$ plane for the same
$\Delta_{\tilde b_2}$ is shaded cyanly in Fig \ref{csec}-{\sf
(b)}. The higher lowest $m_{\rm LSP}$ required from the Eq.
(\ref{bsgb}{\sf b}) (see, also, Fig \ref{5sb}-{\sf (d)}) prevents
the further increase of $\sigma_{\tilde\chi p}^{\rm SI}$.

We observe, also, that the widths of the corresponding bands
(especially of the grey one) are narrower in this case than in the
others. This can be explained by the following observation: In the
present case the major contribution to the $\sigma_{\tilde\chi
p}^{\rm SI}$ calculation comes from $B_D$ and $B_{1D}$ which are
proportional to the hadronic input
$f^{p}_{_{TG}}=1-\sum_{u,d,s}f^{p}_{_{T_q}}$. Varying the inputs
of Eq. (\ref{rgis}) within their ranges, $f^{p}_{_{TG}}$ varies by
almost $15\%$. On the contrary, in the other cases, the major
contribution to the $\sigma_{\tilde\chi p}^{\rm SI}$ comes from
$f_q^{(\tilde q)}$ for $q=s$ (Eq. 40 of Ref. \cite{drees1}) which
is multiplied by $f^{p}_{_{T_s}}$. This input varies by 70$\%$,
and so, it produces a much more wide band on the $m_{\rm
LSP}-\sigma_{\tilde\chi p}^{\rm SI}$ plane. In the case of the
cyan band, the contribution from $f_q^{(\tilde q)}$ for $q=s$
becomes eventually sizable and so, the band is somehow widen for
larger $m_{\rm LSP}$.

Note that if we had imposed UGMs with $\mu<0$ and $r_{\tilde
f}=0.4$, the lower bound on $m_{\rm LSP}$ would have been derived
from Eq. (\ref{g2t}{\sf a}) with result $m_{\rm
LSP}\gtrsim149.5~{\rm GeV}$. Consequently, a region similar to
this in Fig. \ref{5sb}-{\sf (b)} would have been allowed with
maximal $\Delta_{\tilde b_2}\simeq0.13$.

\newpage

\begin{figure}[t]
\hspace*{-.71in}
\begin{minipage}{8in}
\epsfig{file=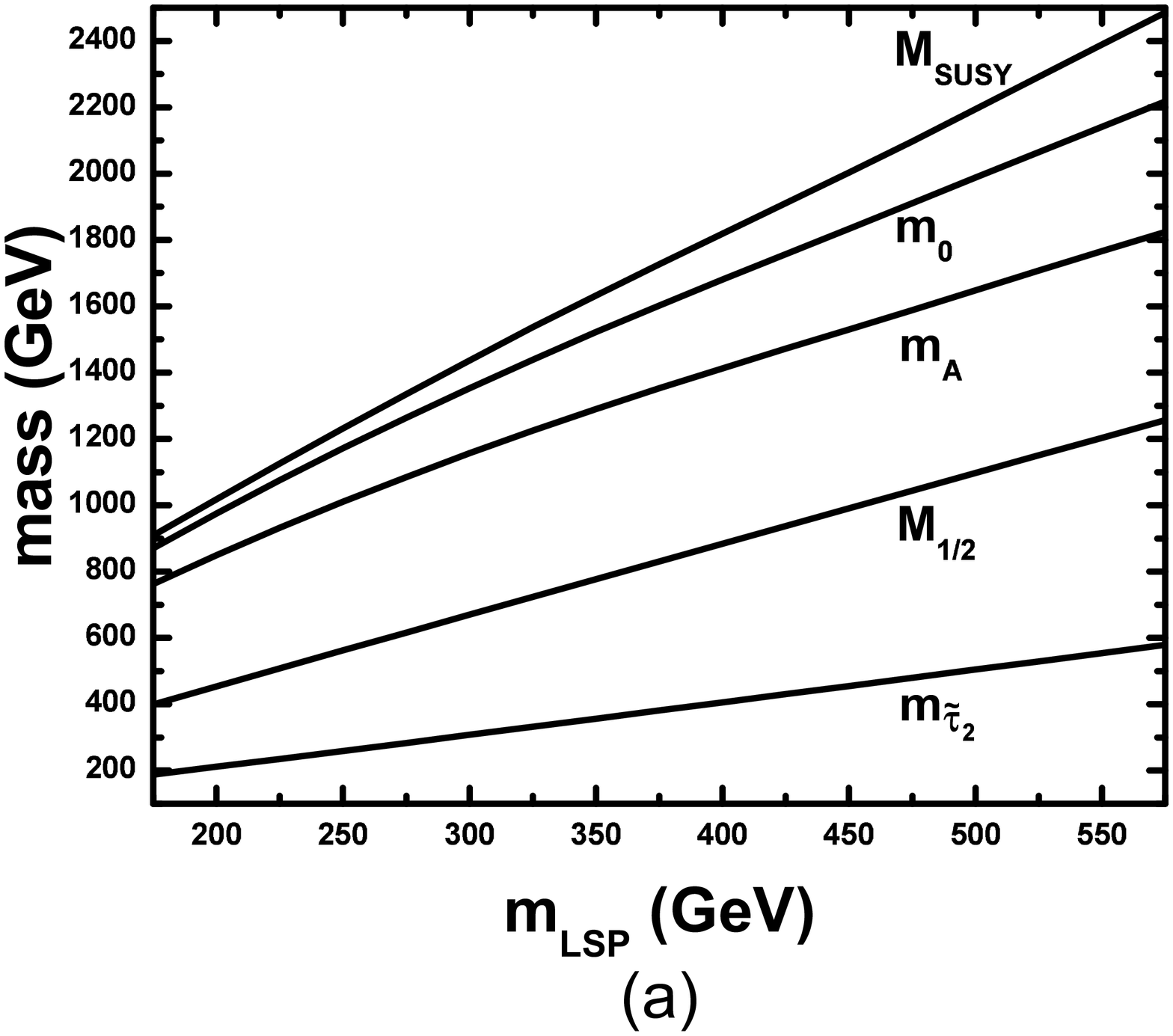,height=3.8in,angle=-90} \hspace*{-1.37 cm}
\epsfig{file=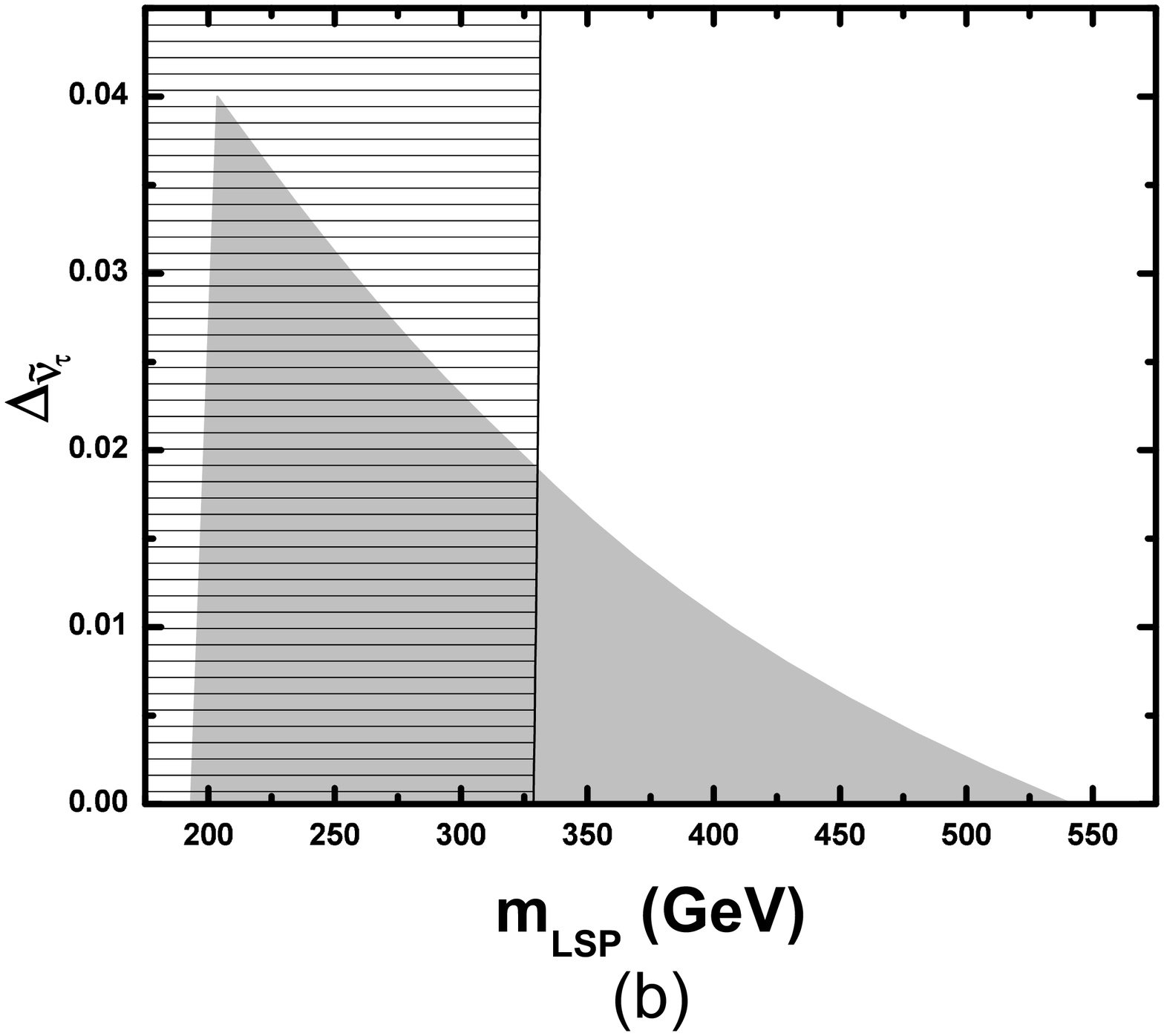,height=3.8in,angle=-90} \hfill
\end{minipage}
\hfill\hspace*{-.71in}
\begin{minipage}{8in}
\epsfig{file=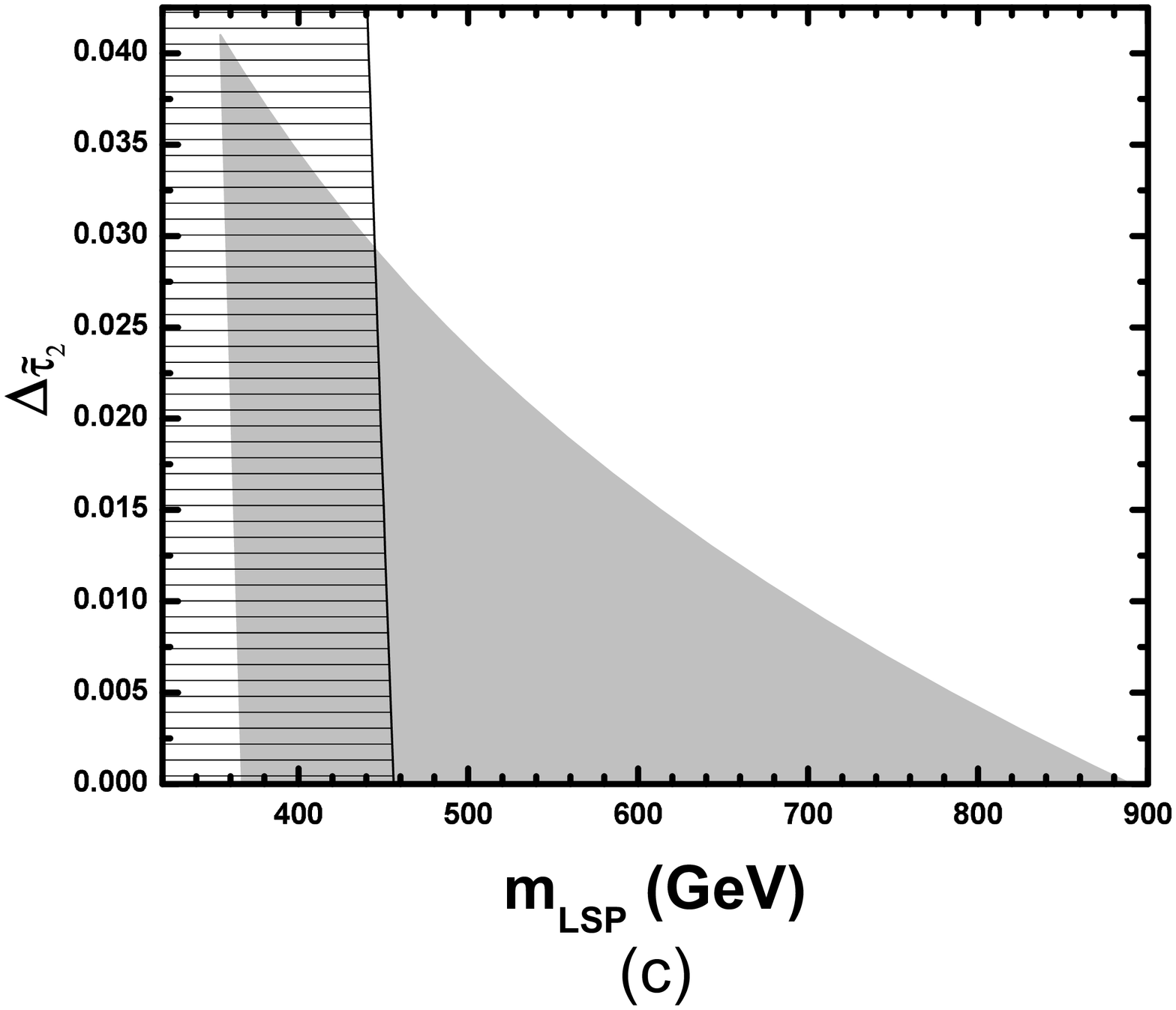,height=3.8in,angle=-90} \hspace*{-1.37 cm}
\epsfig{file=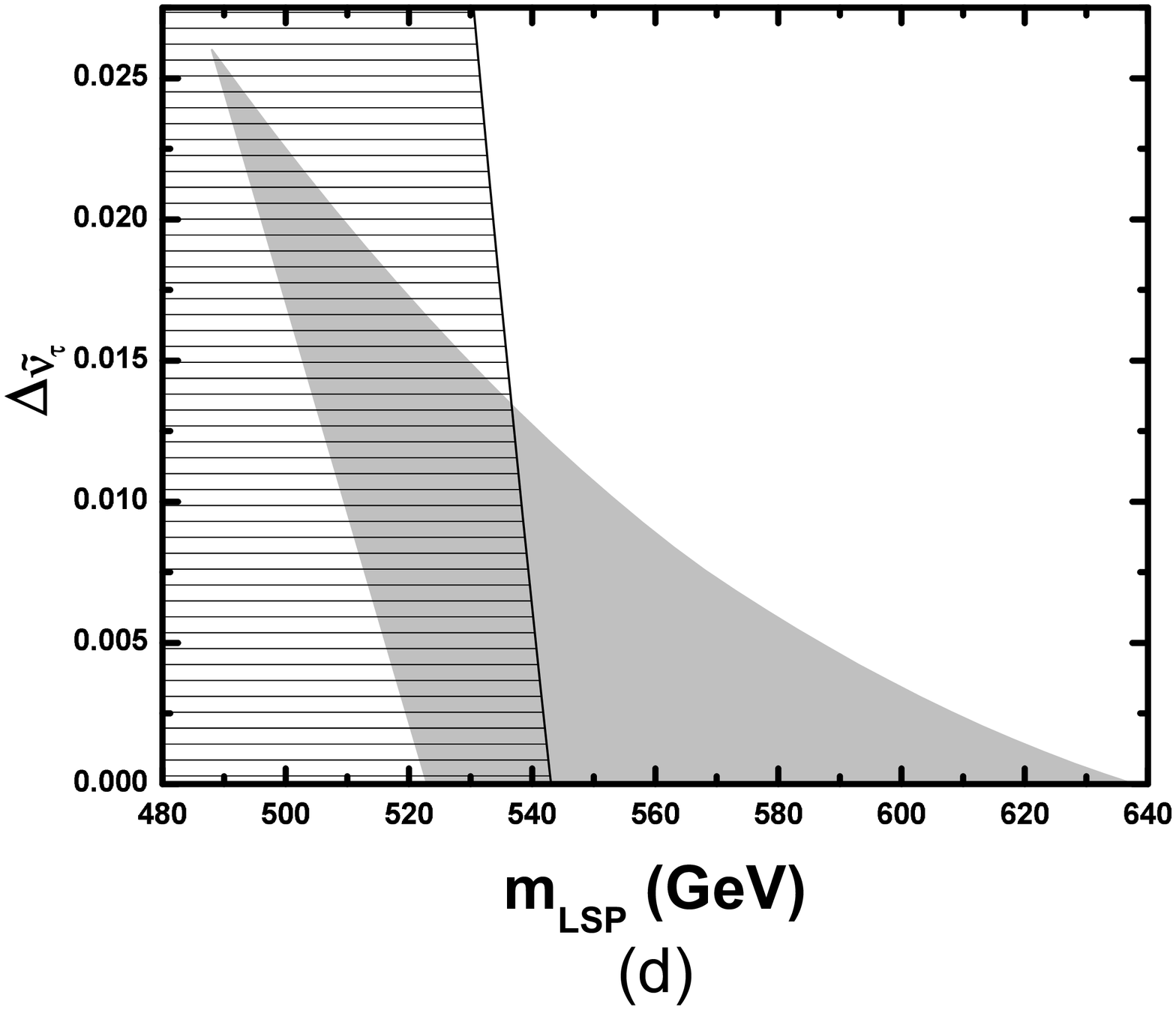,height=3.8in,angle=-90} \hfill
\end{minipage}
\hfill \caption[]{\sl The mass parameters $M_{1/2}$, $m_0$, $m_A$,
$M_{\rm SUSY}$ and $m_{\tilde\tau_2}$ versus $m_{\rm LSP}$ for
$r_{\tilde f}=0.2, r_2=r_3=1$ and $\Delta_{\tilde\nu_\tau}=0$ {\sf
(a)} and the allowed (shaded) area for $r_{\tilde f}=0.2~[0.4],
r_2=1~[0.6], r_3=1~[0.6]$ {\sf (b [d])}, on the $m_{\rm
LSP}-\Delta_{\tilde\nu_\tau}$ plane and for $r_{\tilde f}=0.4,
r_2=0.6, r_3=1$ on the $m_{\rm LSP}-\Delta_{\tilde\tau_2}$ plane
{\sf (c)}. Ruled is, also, the area favored by the optimistic
upper bound on $m_{\rm LSP}$ from Eq. (\ref{g2e}{\sf
a}).}\label{5neu}
\end{figure}

\subsection{$\nta(-\chb-\ntb)-\ntn-\stau$ C{\ssz OANNIHILATIONS}}
\label{resultn}

\hspace{.562cm} For low enough $r_{\tilde f}$'s and $r_2=r_3=1$,
to keep $\tilde b _2$ heavier than $\tilde\chi$, we must decrease
the difference between the values $M_{1/2}$ and $m_0$ depicted in
Fig. \ref{5sb}-(a). The increase of $M_{1/2}$ increases
efficiently $m_{\tilde Q}$ and $m_{\tilde b^c}$ whose the running
(see e.g. Ref. \cite{rge}) depends crucially on $M_3$ but does not
affect a lot $m_{\tilde L}$, which is anyway low at $M_{\rm GUT}$,
due to Eq. (\ref{de}). Therefore, an even new type of CAM between
$\nta-\ntn-\stau$ emerges for $0<r_{\tilde f}<0.27$. Note that the
sneutrino of the two first generations do not participate to this
phenomenon remaining heavier, since their running does not depend
on $m_{\tilde L}$.

We present the mass parameters, in this case, $M_{1/2}$, $m_0$,
$m_A$ and $M_{\rm SUSY}$ together with $m_{\tilde\tau_2}$ versus
$m_{\rm LSP}$ for $\Delta_{\tilde\nu_\tau}=0$, $r_{\tilde f}=0.2$
and $r_2=r_3=1$ in Fig. \ref{5neu}-{\sf (a)}. We see that
$m_{\tilde\tau_2}$ is just 6\% heavier than $m_{\rm LSP}$ and
consequently, participates to the CAM (unlike the similar case of
Ref. \cite{ellis3}). A typical example of the resulting allowed
area on the $m_{\rm LSP}-\Delta_{\tilde\nu_\tau}$ plane is
displayed in Fig. \ref{5neu}-{\sf (b)} for the same $r_2$ and
$r_3$. The left (almost vertical) [right curved] boundary of the
allowed (shaded) region comes from Eq. (\ref{bsgb}{\sf
b}~[\ref{cdmb}]). Similar is, also,  the origin of the boundaries
of the allowed regions for $r_{\tilde f}=0.4, r_2=0.6$ and
$r_3=1~[0.6]$ presented in Fig. \ref{5neu}-{\sf (c [d])}. There,
we fix $r_{\tilde f}=0.4$ since for lower $r_{\tilde f}$,
$\tilde\nu_\tau$ turns out to be lighter than $\tilde\chi$. Also,
we take $r_3=0.6$, and not 0.5 as previously, since for lower
$r_3$'s, $\tilde b _2$ becomes predominantly coannihilator. In the
case of Fig. \ref{5neu}-{\sf (c)}, $m_{\tilde\tau_2}$ turns out to
be slightly lighter than $m_{\tilde\nu_\tau}$ and so, the allowed
area is presented on the $m_{\rm LSP}-\Delta_{\tilde\tau_2}$
plane. Also, the allowed region of Fig. \ref{5neu}-{\sf (d)} is
disconnected to this of Fig. \ref{5pb}-{\sf (b)}.

\begin{table}[!t]
\caption {\bf D\ssz OMINANT \nsz C\ssz ONTRIBUTIONS TO \nsz
{\boldmath $\Omega_{\rm LSP}h^2$} }
\begin{center}
\begin{tabular}{|l||c|c|c|}\hline
\multicolumn{4}{|c|}{\bf M\ssz ODEL \nsz P\ssz ARAMETERS} \\
\hline
{$r_{\tilde f}$}&{$0.2$}&{$0.4$}&{$0.4$}\\
{$r_2, r_3$}&{$r_2=r_3=1$}&{$r_2=0.6, r_3=1$}&{$r_2=0.6,
r_3=0.6$}\\ \hline \hline
\multicolumn{4}{|c|}{\bf A\ssz LLOWED \nsz R\ssz ANGES}\\ \hline
$\tan\beta$&{$34.2-37.7$}&{$34.3-36.6$}&{$34.8-35.5$} \\
$m_{\rm LSP}~({\rm GeV})$ &{$201-(330)542$}
&{$354-(456)890$}&{$488-(543)638$}\\
$\Delta_{\tilde\nu_\tau}$&{$0.04-0$}&{$0.055-0.002$}&{$0.026-0$}\\
\hline \hline
\multicolumn{4}{|c|}{\bf P\ssz ROCESSES \nsz W\ssz HICH \nsz C\ssz
ONTRIBUTE \nsz M\ssz ORE \nsz T\ssz HAN \nsz {\boldmath $5\%$}}
\\ \hline \multicolumn{1}{|l||}{\bf P\ssz ROCESS} &
\multicolumn{3}{|c|}{\bf C\ssz ONTRIBUTION \nsz ({\boldmath
$\%$})}\\ \cline{2-4} \hline
$\nta\nta\rightarrow b\bar b$ &$-$&$12-6$&$16-1$\\ \hline
$\nta\ntn\rightarrow \tau W^-$ &$20-6$&$7-3$&$4-4$\\
$ \nta\ntn\rightarrow \nu_\tau Z $ &$11.6-0$&$-$&$-$\\\hline
$\nta\stau\rightarrow \tau h$ &$-$&$6-1.5$&$-$\\
$\nta\stau\rightarrow \nu_\tau W^-$ &$-$&$7-1$&$6-4$\\
$ \nta\stau\rightarrow \tau Z $ &$-$&$6-2$&$-$\\\hline
$\ntn\ntn\rightarrow \nu_\tau\nu_\tau$ &$19-17$&$-$&$5-12$\\
$\ntn\ntns\rightarrow ZZ $&$10.5-12$&$-$&$3-7$\\
$\ntn\ntns\rightarrow W^+W^-$ &$12-13.5$&$-$&$3-7$\\\hline
$\stau\ntns\rightarrow \nu_\tau\tau$&$0-9$&$-$&$-$\\
$\stau\ntns\rightarrow \gamma W^-$&$2-6$&$-$&$-$\\
$\stau\ntns\rightarrow \bar t b$&$-$&$-$&$2-18$\\
$\stau\ntn\rightarrow \nu_\tau\tau$ &$0.5-2.6$&$-$&$5-10$\\\hline
$\stau\stau\rightarrow \tau\tau$ &$0.7-6$&$5-6$&$4-10$\\
$\stau\staus \rightarrow W^+W^-$ &$0.6-6$&$-$&$3-7$\\
$\stau\staus \rightarrow b \bar b$ &$-$&$3-20$&$-$
%
\\ \hline
\end{tabular}
\end{center}
\end{table}

The allowed ranges of $m_{\rm LSP}$, $\Delta_{\tilde \nu_\tau}$
and $\tan\beta$ are listed in the Table 3. For $m_{\rm LSP}$
indicated in the parenthesis, satisfaction of Eq. (\ref{g2e}{\sf
a}) occurs, too. We deduce that, despite the presence of an extra
coannihilator, the reduction of $\Omega_{\rm LSP}h^2$ caused by
$\nta-\ntn-\stau$ CAM is not more efficient than the case
$\nta-\stau$ CAM and therefore, the maximal allowed $m_{\rm
LSP}$'s do not essentially differ. However, thanks to the heavier
higgs sector, Eq. (\ref{bsgb}{\sf b}) is satisfied for lighter
charginos and neutralinos, and consequently (see sec.
\ref{phenod}), there is parameter space, where the putative bound
of Eq. (\ref{g2e}{\sf a}) can be fulfilled, too. We observe, also,
that although the contribution of $\nta-\chb-\ntb$ CAM for the
used $r_2=r_3\neq1$ is very weak, it leads to a sizable
suppression of $\Omega_{\rm LSP}h^2$ increasing the upper bound on
$m_{\rm LSP}$.

The $\sigma_{\tilde\chi p}^{\rm SI}$ versus $m_{\rm LSP}$  for
$\Delta_{\tilde \nu_\tau}=0$, $r_{\tilde f}=0.2$ and $r_2=r_3=1$
is depicted in Fig. \ref{csec}-{\sf (b)}, light grey band. We
observe that although the maximal $\sigma_{\tilde\chi p}^{\rm SI}$
lies below the range of Eq. (\ref{sgmb}), it is significantly
higher than in the case with $\Delta_{\tilde\tau_2}=0$ and
$r_{\tilde f}=r_2=r_3=1$ (dark grey band), due to the easier
satisfaction of Eq. (\ref{bsgb}{\sf b}). Possible diminution of
$r_2$ and/or $r_3$ do not produce any improvement, since the
lowest possible $m_{\rm LSP}$, derived again from Eq.
(\ref{bsgb}{\sf b}), increases enough (note that ${\rm BR(b\to
s\gamma)}|_{\rm SUSY}$ increases lowering $m_{\tilde\chi^\pm}$).

If we had assumed UGMs with $\mu<0$ and $r_{\tilde f}=0.2$, the
lower bound on $m_{\rm LSP}$ would have been derived from Eq.
(\ref{g2t}{\sf a}) with result $m_{\rm LSP}\gtrsim289~{\rm GeV}$.
Consequently, a region similar to this of Fig. \ref{5neu}-{\sf
(b)} would have been allowed with maximal
$\Delta_{\tilde\nu_\tau}\simeq0.025$.


\section{C{\ftn ONCLUSIONS}-O{\ftn PEN} I{\ftn SSUES}}\label{con}

\hspace{.562cm} We considered a MSSM version which could emerge
from the breakdown of a $SU(5)$ or $SO(10)$ SUSY model at GUT
scale. Namely, we assumed $b-\tau$ YU, allowing gaugino and
sfermion mass non-universality. We then restricted the parameter
space of the model by imposing the constraints from CDM, SUSY
corrections to $b$-quark mass, ${\rm BR}(b \rightarrow s\gamma)$,
$\delta\alpha_\mu$ and accelerator data and derived scalar
neutralino-proton cross sections. SUSY spectra and scalar
neutralino-proton cross sections were calculated by using our
numerical code, while the values of the various constraints, by
employing the current version of {\tt micrOMEGAs} package,
supplemented by an updated ${\rm BR}(b\rightarrow s\gamma)$ code.

We showed that an opposite sign on the asymptotic gluino mass,
based on group theoretical grounds, assists us to succeed
compatibility between the $b-\tau$ YU and the lower bound from the
$\delta\alpha_\mu$ constraint ($e^+e^-$-based calculation). We,
then, parameterized the possible non-universality in the (i)
gaugino sector, defining $r_2,~ r_3$ as the ratios between the
asymptotic wino and gluino masses with the bino (ii) sfermion
sector, defining $r_{\tilde f}$, as the ratio between the
asymptotic sfermion masses in ${\bf 10}$ and ${\bf \bar5}$ reps of
$SU(5)$ (iii) Higgs sector, defining $r_{H_1},~r_{H_2}$  as the
ratio between the two asymptotic Higgs masses and asymptotic
sfermion masses in ${\bf 10}$ reps. We used universal Higgs masses
with $r_{H_1}=r_{H_2}=1$.

We found regions of the parameter space consistent with all the
imposed restrictions, paying special attention to each applied
$\Omega_{\rm LSP}h^2$ reduction ``procedure''. Regarding this
issue for $r_2=r_3=1$, we can distinguish the cases: (i) For
$0<r_{\tilde f}\leq1.2~[\sqrt{2}]$, $A$PE [and/or $\nta-\stau$
CAM] can drastically reduce $\Omega_{\rm LSP}h^2$ and succeed to
bring it below the CDM upper bound for $m_{\rm LSP}$'s allowed by
${\rm BR}(b\rightarrow s\gamma)$. The LSP mass can be as low as
$205~{\rm GeV}$ for $r_{\tilde f}=1$. (ii) For $0.27\leq r_{\tilde
f}\leq0.56$, $\nta-\sbb$ CAM can be activated. The lowest possible
LSP is $83~{\rm GeV}$ for $r_{\tilde f}=0.48$ but much heavier
residual SUSY spectrum. (iii) For $0<r_{\tilde f}<0.27$,
$\nta-\ntn-\stau$ CAM can be applied. The lowest possible LSP is
$174~{\rm GeV}$ for $r_{\tilde f}=0.26$ and not too heavier SUSY
spectrum. In both latter cases, satisfaction of the optimistic
upper bound from $\delta\alpha_\mu$ can be, also, achieved in
sharp contrast with the universal-like case (i). Interesting
scalar neutralino-proton cross section is obtained in the case of
$\nta-\sbb$ CAM due to the $\nta-\sbb$ proximity and the light
lowest $m_{\rm LSP}$. If we had imposed UGMs, we would have had
qualitatively similar results, with, in general, higher
$\tan\beta$'s and lowest $m_{\rm LSP}$'s but with restrictions on
the parameters, derived from the $\tau$-based calculation of
$\alpha^{\rm SM}_\mu$.

In our investigation, we considered, also, cases for $r_2=0.6$
and/or $r_3=1, 0.6, 0.5$. In these, gaugino inspired CAMs
($\nta-\chb-\ntb$) can be activated and combined with the former
sfermionic CAMs, creating new, privileged situations in the
$\Omega_{\rm LSP}h^2$ calculation without to reduce it far lower
than the expectations. In most cases the lowest and the highest
possible $m_{\rm LSP}$'s are higher than the former cases.
Consequently, no important improvement on the maximal scalar
neutralino-proton cross section is observed, although the gaugino
purity of LSP is decreased.

Lastly, we should discuss the fate of the sfermionic CAMs in the
predictive cases in which the arrangement of Eq. (\ref{ratio}) is
``spontaneously'' produced. Namely with dominance of (i) {\bf 24}
reps  in Eq. (\ref{genr}) for the case of $SU(5)$, we take
\cite{eent} $r_2=3$ and $r_3=2$. We observe that both new
sfermionic CAMs can be activated for lower $r_{\tilde f}$ 's. (ii)
{\bf 54} reps in Eq. (\ref{genr}) for the case of $SO(10)$, and
for the symmetry breaking pattern $SO(10)\to SU(4)_{\rm c}\otimes
SU(2)_L\otimes SU(2)_R$, we take \cite{so10g} $r_2=3/2$ and
$r_3=1$ with results quite similar to those with $r_2=r_3=1$ (iii)
{\bf 54} reps in Eq. (\ref{genr}) for the case of $SO(10)$ and for
the symmetry breaking pattern $SO(10)\to SU(2)\otimes SO(7)$, we
take $r_2=-7/3 $ and $r_3=-1$. $\nta-\sbb$ CAM is similarly
achieved but $\nta-\ntn-\stau$ CAM can not occur.

Our results do not crucially dependent on our choice
$r_{H_1}=r_{H_2}=1$. We checked that if we had imposed
$r_{H_1}=r_{H_2}=r_{\tilde f}$, we would have obtained similar
sfermionic CAMs but for much lower $r_{\tilde f}$ 's than those
used here. The situation with $r_{H_1}\neq r_{H_2}$ requires
certainly deeper investigation, since, then, additional
contributions in the RGEs arise \cite{ellis3} and much more rich
situations can emerge.

For simplicity, $m^{\rm c}_b(M_Z)$ was fixed to its central
experimental value throughout our calculation. Allowing it to vary
within its $95\%$ c.l. range of Eq. (\ref{mbb}), the allowed
ranges of $\tan\beta$ will further widen. At least, the
non-universality in the gaugino sector can cause, through $c_i$ of
Eq. (\ref{genr}), interesting consequences to the GUT structure of
the theory, such as to the proton stability (see, e.g. Refs.
\cite{murayama, kawam, pati}). However, such an analysis is
outside the scope of this work.


\acknowledgments \hspace{.562cm} The author is grateful to {\tt
micrOMEGAs} team, G. B\'{e}langer, F. Boudjema, A. Pukhov and A.
Semenov, for providing their updated ${\rm BR}(b\rightarrow
s\gamma)$ code. He, also, wishes to thank S. Bertolini and A.
Masiero for useful discussions, B. Bajc, R. Derm\'{\i}\v sek and
S. Khalil for interesting comments. Collaboration in the early
stage of this work with S. Profumo  is acknowledged, too. This
research was supported by European Union under the RTN contract
HPRN-CT-2000-00152.

\end{document}